\documentclass[11pt,a4paper]{article}
\pdfoutput=1
\usepackage{amssymb}
\usepackage{amsmath}
\usepackage{mathtools}
\usepackage{subcaption}
\usepackage{ physics }
\usepackage[compat=1.1.0]{ tikz-feynman }
\usepackage{ tikz-feynhand }
\usepackage{ array }
\usepackage{ tensor }
\usepackage{ bbm }
\usepackage{youngtab}
\usepackage{amsfonts}
\usepackage{float	}
\usepackage[font=footnotesize, justification=centering]{caption}
\graphicspath{ {figs/} }
\usepackage{jheppub}
\usepackage[export]{adjustbox}
\usepackage{slashed}
\usepackage{IEEEtrantools}
\usepackage{tabu}
\usepackage{longtable}
\IEEEeqnarraydefcolsep{0}{\leftmargini}
\usepackage{lineno}
\usepackage{rotating}
\usepackage{amsmath,amssymb,graphicx,slashed,xcolor,cancel}
\usepackage{rotating}
\usepackage{array,booktabs,colortbl,multirow,tablefootnote}
\newcommand{\vtext}[1]{\begin{sideways}{#1}\end{sideways}}
\usepackage[normalem]{ulem}
\usepackage{color,colordvi}

\newcommand{\dlo}[1]{\ensuremath{\partial_{#1}}}
\newcommand{\dup}[1]{\ensuremath{\partial^{#1}}}
\newcolumntype{C}{>{\centering\arraybackslash}b{1.5cm}}
\renewcommand{\epsilon}{\varepsilon}
\newcommand{\be}{\begin{equation}}
\newcommand{\ee}{\end{equation}}
\newcommand{\bead}{\begin{aligned}}
\newcommand{\eead}{\end{aligned}}
\newcommand{\bmat}{\left(\begin{matrix}}
\newcommand{\emat}{\end{matrix}\right)}

\newcommand{\epsl}[1]{\ensuremath{\varepsilon_{#1}}}

\toccontinuoustrue
\abstract{
	From general analyticity and unitarity requirements on the UV theory, positivity bounds on the Wilson coefficients of the dimension-8 operators composed of 4 fermions and two derivatives appearing in the Standard Model Effective Field Theory have been derived recently. We explore the fate of these bounds in the context of models endowed with a Minimal Flavor Violation (MFV) structure, models in which the flavor structure of higher dimensional operators is inherited  from the one already contained in the Yukawa sector of the Standard Model Lagrangian. Our goal is to check whether the general positivity bounds translate onto bounds on the Yukawa coefficients and/or on elements of the CKM matrix. MFV fixes the coefficients of dimension-8 operators up to some multiplicative flavor-blind factors and we find that, in the most generic setup, the freedom left by those unspecified coefficients is enough as not to constrain the parameters of the renormalizable Yukawa sector. On the contrary, the latter shape the allowed region for the former. Requiring said overall coefficients to take natural $\order{1}$ values could give rise to bounds on the Yukawa couplings. Remarkably, at leading order in an expansion in powers of the Yukawa matrices, no bounds on the CKM entries can be retrieved.
}

\preprint{DESY 20-201}

\title{Positivity bounds on Minimal Flavor Violation}

\author[a]{Quentin Bonnefoy,}
\author[a,b]{Emanuele Gendy,}
\author[a,c]{Christophe Grojean}

\affiliation[a]{Deutsches Elektronen-Synchrotron (DESY), D-22607 Hamburg, Germany}
\affiliation[b]{Institute of Theoretical Physics, Universit\"at Hamburg, 22761 Hamburg, Germany}
\affiliation[c]{Institut f{\"u}r Physik, Humboldt-Universit{\"a}t zu Berlin, D-12489 Berlin, Germany} 

\emailAdd{quentin.bonnefoy@desy.de}
\emailAdd{emanuele.gendy@desy.de}
\emailAdd{christophe.grojean@desy.de}

\begin{document}
	\maketitle
	\flushbottom
	
%%%%%%%%%%%%%%%%%%%%%%%%%%%%%%%%%%%%%%%%%%%%%%%%%%%%%%%%%%%%%%%%%%%%%%%%%%%%%%%%%%%%%%%
\section{Introduction}

Effective Field Theories (EFTs) are one of the most powerful and general ways to describe a physical system, when one does not know or care about its dynamics below some length scale $\Lambda^{-1}$. In fact, unless one is addressing the final theory of everything, this is always the case. The procedure behind EFTs consists in writing down the most general, non-renormalizable Lagrangian as a series of gauge invariant operators built out of the relevant degrees of freedom, and suppressed by appropriate inverse powers of $ \Lambda$. Then, as long as one is working at masses and exchanged momenta $m,|\vec{p}|\ll \Lambda$, such Lagrangian can be employed to make predictions by retaining the relevant terms in the expansion. The Standard Model (SM) itself has to be considered as an effective description of Nature, and indeed it corresponds to the leading order, low energy approximation of the more complete Standard Model Effective Field Theory (SMEFT). The latter is defined as the EFT where the SM degrees of freedom are employed to build all the operators invariant under the SM gauge group, and takes the general form:
	\begin{align}
	\mathcal{L}_{SMEFT}=\mathcal{L}^{(4)}_{SM}+\sum_{n\geq 5}\frac{c_n}{ \Lambda^{n-4}}\mathcal{O}^{(n)} \ .
	\end{align}
	Here $\mathcal{L}^{(4)}_{SM}$ is the dimension-4, renormalizable part of the Lagrangian, $ \Lambda$ is a UV scale until which the SMEFT is valid and above which its predictivity is lost, and $\mathcal{O}^{(n)}$ are gauge-invariant, dimension-$n$ operators.  The $c_n$ are coefficients which, after taking into account $\hbar$ dimensions, and possible selection rules (for instance baryon number conservation), are to be taken in principle to be of $\order{1}$.
	
	However, this is not the end of it. Indeed, not all the apparently healthy EFTs are actually consistent. Instead, requiring that their UV completion respects unitarity and analyticity, two properties that we wish upon any Quantum Field Theory (QFT), imposes bounds on some of the Wilson coefficients entering the EFT Lagrangian~\cite{irobstruct}. In particular, the simplest bounds arise from positivity requirements on the $s^2$ coefficient of the IR $2\to2$ amplitude, with $s$ the Mandelstam variable $s=(p_1+p_2)^2$.
	This property has had a variety of applications, from pion physics~\cite{pions1,pions2} to Quantum Gravity~\cite{qg} and to the derivation of the $a$-theorem~\cite{atheorem}, see also~\cite{softness} and references therein. Recently, efforts have been made to constrain the coefficients in the SMEFT expansion~\cite{sumrules,Gu:2020thj}, particularly regarding vector boson scattering~\cite{SMEFTconsistency, vectorbosonscattering, aQGC} (see also Refs.~\cite{cones,electronpositronscattering,Yamashita:2020gtt,dimension6bounds,Bellazzini:2020cot,Tolley:2020gtv,Gu:2020ldn,Trott:2020ebl} for recent developments). 
		
	Finally, Ref.~\cite{flavorconstr} obtained results about the coefficients of dimension-8 operators composed of 4 fermions and two derivatives appearing in the SMEFT. There, interesting bounds are derived, for instance for the flavor violating coefficients, whose magnitude turns out to be bounded from above by products of flavor conserving ones. Then, it is natural to check whether these bounds are compatible with the Minimal Flavor Violation (MFV) ansatz~\cite{MFV, MFVreview}. The latter is one of the simplest methods of constraining the flavor structure of any higher dimensional operator in the SM effective theory containing fermions, in a way that does not clash with the stringent experimental bounds on flavor violation. It does this by stating that all the relevant building blocks are already contained in the Yukawa sector of the SM. Thus, the flavor structure of any operator involving fermions is fixed by the dimension-4 ones, up to some overall multiplicative factors. This applies in particular to four-fermions dimension-8 operators, whose coefficients are constrained by positivity bounds. Consequently, assuming MFV, one can ask whether the positivity bounds yield constraints on the parameters of the dimension-4 Lagrangian. For instance, flavor-violating dimension-8 couplings are proportional to entries of the CKM matrix in MFV, and positivity constraints will involve both dimension-4 and dimension-8 coefficients. Therefore, our initial goal consists in checking whether one can extract from the EFT consistency some bounds on the Yukawa and CKM parameters of the SM, to be then compared with experimental value. 
		
	It is a known fact that, whatever the values of the SM parameters, there exist MFV-compatible UV-completions of the SMEFT four-Fermi operators, an example being a heavy vector coupled in a flavor-blind way. So the positivity bounds cannot be as powerful enough as to make some of values of the SM parameters inconsistent.
	 Nonetheless, the positivity bounds will restrict the allowed region for the unspecified EFT coefficients that the MFV assumption already reduced down to flavor-blind global factors. And, the specific values of  the renormalizable couplings influence the shape of the allowed region of the flavor-blind MFV overall factors. Remarkably, while this applies to fermion masses, the structure of the positivity bounds is such that the CKM matrix elements completely disappear from the most stringent bounds at leading order. 
			
		To go further, we consider the expectation that the allowed region for the EFT coefficients should enable the flavor-bling MFV factors to be order one. Otherwise, MFV would be cornered by the EFT consistency to unnatural realizations (or specific ones, like the aforementioned case of a flavor-blind heavy vector), which would question its use in the first place. Consequently, we study whether the assumption of order one EFT coefficients now yields interesting constraints on fermion masses and CKM elements, when implemented in the bounds involving dimension-4 and dimension-8 coefficients. We study this case analytically when all the coefficients are degenerate and equal to one, and numerically when they vary independently in a neighborhood of unity and find indeed a bound on the largest fermion mass. In the present case, those bounds are ineffective, phenomenologically speaking as they are by far satisfied by experimental values, and theoretically speaking since the values that violate them lie in a region where the MFV expansion breaks down, preventing us from obtaining any relevant bound on fermion masses. Nevertheless, they show how in some cases restrictions on higher dimensional operators can get reflected on renormalizable parameters.

	The paper is organized as follows. In Section~\ref{sec:MFV}, we provide a brief review of MFV, and express the coefficients of dimension 8, 4-fermions operators in Section~\ref{sec:MFVansatzSection}, accordingly. In Section~\ref{sec:studyingthebounds}, we turn to study the bounds, to remove their dependence on external states, and to find the allowed regions for the global factors that were left undetermined under the MFV assumption. Section~\ref{sec:conclusions} is devoted to final comments and conclusions. Several appendices complete this paper. Appendix~\ref{sec:grouptheoryLargeTop} describes the flavor structures that arise when resumming the MFV expansion with large Yukawas, and Appendix~\ref{sec:grouptheoryrho4} shows that a flavor structure introduced in Section~\ref{sec:MFVansatzSection} is redundant when there are only two flavors. Then, Appendix~\ref{sec:dimension6operators} discusses the impact on the positivity bounds of the SMEFT operators that we neglected. The general bounds in the three flavor case are presented in Appendix~\ref{boundsGeneral} and finally, Appendix~\ref{sec:rhodependence} displays some coefficient redefinitions that we use throughout the paper to simplify the expression of the positivity bounds.

%%%%%%%%%%%%%%%%%%%%%%%%%%%%%%%%%%%%%%%%%%%%%%%%%%%%%%%%%%%%%%%%%%%%%%%%%%%%%%%%%%%%%%%	
\section{Minimal Flavor Violation}
	\label{sec:MFV}
	Minimal Flavor Violation (MFV)~\cite{MFV, MFVreview} is an ansatz constraining the flavor structure of higher dimensional operators in the Standard Model Effective Field Theory (SMEFT). It comes from noticing that, if we do not consider the Yukawa operators, the Lagrangian of the Standard Model enjoys a $U(3)^5$ global symmetry, that acts on the quark and lepton flavor space. Its non-abelian subgroup can be split as:
	\begin{align}
	SU(3)^3_q&=SU(3)_Q\otimes SU(3)_u \otimes SU(3)_d\nonumber\\
	SU(3)^2_l&=SU(3)_L\otimes SU(3)_e,
	\end{align}
	which means that the flavor vector $i\in \{Q,u,d,L,e\}$\footnote{In this work we omit $L$ and $R$ chirality subscripts, and indicate with lowercase $u$, $d$ and $e$ the right-handed up, down quark and electron respectively, and with uppercase $Q$ and $L$ the quark and lepton doublets.} transforms as a fundamental of $SU(3)_i$.
	Then the Yukawa sector of the Lagrangian
	\begin{align}
	\mathcal{L}_{\text{Yukawa}}=\bar{Q}Y_ddH + \bar{Q}Y_uu\tilde{H} + \bar{L}Y_eeH+\text{ h.c.}
	\end{align}
can be made formally invariant under this group if we promote the Yukawa matrices, $Y_{u,d,e}$, to spurion fields transforming as in Table~\ref{tab:ytrasmforma}. 
	\begin{table}[H]
		\centering
		\begin{tabular}{C|C|C|C|C|C}
			& $SU(3)_Q$ & $SU(3)_u$ & $SU(3)_d$ & $SU(3)_L$ & $SU(3)_e$\\\hline\vspace{0.1cm}
			$Y_u$ &$\mathbf{3}$ & $ \mathbf{\bar{3}}$ & $\mathbf{1}$ &$\mathbf{1}$ &$\mathbf{1}$  \\[0.1cm]
			$Y_d$ &$\mathbf{3}$ & $ \mathbf{1}$ & $\mathbf{\bar{3}}$ &$\mathbf{1}$ &$\mathbf{1}$  \\[0.1cm]
			$Y_e$ &$\mathbf{1}$ & $ \mathbf{1}$ & $\mathbf{1}$ &$\mathbf{3}$ &$\mathbf{\bar{3}}$
		\end{tabular}
		\captionsetup{width=.75\textwidth}
		\caption{Transformation properties of the Yukawa matrices treated as spurions under the MFV assumption.}
		\label{tab:ytrasmforma}
	\end{table}
Minimal Flavor Violation is the requirement that any higher dimensional operator has to be built out of $Y$ matrices and Standard Model fields, and must be formally invariant under the flavor group, taking into account the transformation properties in Table~\ref{tab:ytrasmforma}. 
	Notice that the spurions transform under the $U(1)$ abelian factors of $U(3)^5$, too. We do not treat those explicitly, but they turn out to be useful, e.g., to get rid of terms like $\sim \big(Y_u\big)^m\big(Y_u^\dagger\big)^n \left(\bar Q Q\right)^p$ with $m\neq n$.
	
	After building the Lagrangian, we can set the spurion fields to their vacuum values, namely the physical Yukawas. The latter are defined up to the freedom of changing the fermion fields basis, i.e., they are defined up to $U(3)^3_q\otimes U(3)^2_l$ transformations. In the following, we choose a basis where they read:
	\begin{align}
	Y_u&=\lambda_u &Y_d&=V_\textit{CKM}\lambda_d & Y_e&=\lambda_e \ ,
	\label{eq:interactionbasis}
	\end{align}
	where the $\lambda$'s are diagonal matrices containing the diagonal Yukawa couplings, e.g., $\lambda_u=\text{diag}(y_u,y_c,y_t)$, and $V_\textit{CKM}$ is the CKM matrix. Another basis that we will occasionally mention is the following,
	\begin{align}
	Y_u&=V_\textit{CKM}^\dagger\lambda_u &Y_d&=\lambda_d & Y_e&=\lambda_e \ ,
	\label{eq:interactionbasis2}
	\end{align}
related to the previous one via a $U(3)_Q$ transformation.
	As we will make clear in Section~\ref{sec:flavorviolation}, our discussion cannot and does not depend on the particular choice of basis. 
	The MFV framework is relevant and particularly convenient from a theoretical point of view, since it drastically reduces the number of free parameters entering the Lagrangian at each mass-dimension~\cite{MFVandU2}. In addition, its phenomenological value resides in the fact that, in an unconstrained setting, the coefficients of operators contributing to Flavor Changing Neutral Currents (FCNC) would be naturally of $\order{1}$. This is in contradiction with experiments, at least up to New Physics scales of $\sim\order{10^3\,\text{TeV}}$. MFV is an efficient way, albeit perhaps drastic, to justify such behaviour~\cite{MFVreview}.
	
%%%%%%%%%%%%%%%%%%%%%%%%%%%%%%%%%%%%%%%%%%%%%%%%%%%%%%%%%%%%%%%%%%%%%%%%%%%%%%%%%%%%%%%	
	\subsection{Dimension 8 independent fermionic operators}
	
	As shown in Ref.~\cite{irobstruct}, requiring unitarity and analyticity to be properties of the theory up to the UV imposes bounds on some of the coefficients of the Effective Field Theory expansion. What we wish to see is to what degree these bounds are compatible with the MFV assumptions. Thus, the most interesting approach consists perhaps on focussing on 4-fermions operators. However, the lowest order, dimension 6 operators, giving no $s^2$ contribution to the forward amplitude, are unaffected by these bounds (see however Ref.~\cite{dimension6bounds} for recent efforts towards constraining such operators). Thus, as anticipated, we will focus on how these requirements affect the coefficients of operators formed by 4 fermionic fields and two derivatives. Bounds for those coefficients have been obtained in Ref.~\cite{flavorconstr}, whence we will take part of the terminology and conventions adopted in the rest of the paper. In particular, operators formed with fields of one kind only, i.e., those of the schematic form $\mathcal{O}\sim\partial^2(\bar{\psi}_m\Gamma\psi_n)(\bar{\psi}_p\Gamma\psi_q)$, $\psi\in\{u,d,Q\}$, where $\Gamma$ is some combination of Dirac and SM gauge matrices and we only made flavor indices explicit, will be dubbed \textit{self-quartic}. We will refer to those formed with two kinds of fields, $\mathcal{O}\sim\partial^2(\bar{\psi}_m\Gamma\psi_n)(\bar{\chi}_p\Gamma\chi_q)$, $\psi,\chi\in\{u,d,Q\}$ and $\psi\neq\chi$, as \textit{cross-quartic}. The way Lorentz indices are contracted is not shown here.
	We restrict for this discussion to the quark sector only. The extension of our methods to include leptons can then be readily found\footnote{This is true provided one does not include right-handed neutrinos in the discussion. Then, a generalization of MFV accounting for the Pontecorvo--Maki--Nakagawa--Sakata (PMNS) mixing matrix is needed~\cite{MLFV}.}. The list of independent operators we are interested in, then, can be read off Table~\ref{tab:operators}. Operators of the form $	\mathcal{O}=\dlo{\mu}(\bar{\psi}_m\gamma_\nu \psi_n)\dup{\mu}(\bar{\chi}_p\gamma^\nu \chi_q)$, $\psi\neq\chi$ are also present, and are independent from the ones listed in Table~\ref{tab:operators}. However, since they do not contribute to the forward amplitude, there are no bounds on their Wilson coefficients~\cite{flavorconstr}.
	
\begin{table}[ht]
\begin{center}
\begin{tabular}{cclc}
\toprule
Type & Content & \multicolumn{1}{c}{Operator} & Symmetry\\
\midrule
\multirow{8}{*}{\vtext{self-quartic\quad \quad \quad \quad}} & 
% 4u
\multirow{2}{*}{(4-u)} & 
$\mathcal{O}_1[u]=c^{u,1}_{mnpq}\dlo{\mu}(\bar{u}_m\gamma_\nu u_n)\dup{\mu}(\bar{u}_p\gamma^\nu u_q)$ & \\[1mm]
& & $\mathcal{O}_3[u]=c^{u,3}_{mnpq}\dlo{\mu}(\bar{u}_mT^a\gamma_\nu u_n)\dup{\mu}(\bar{u}_pT^a\gamma^\nu u_q)$ &\multirow{6}{*}{\vtext{$c_{mnpq}=c_{pqmn}$\qquad\qquad}\quad\vtext{$c_{mnpq}=c^*_{nmqp}$\qquad\qquad}}\\
\cmidrule{2-3}
%4Q
& \multirow{4}{*}{(4-Q)} & $\mathcal{O}_1[Q]=c^{Q,1}_{mnpq}\dlo{\mu}(\bar{Q}_m\gamma_\nu Q_n)\dup{\mu}(\bar{Q}_p\gamma^\nu Q_q)$ &\\[1mm]
&&$ \mathcal{O}_2[Q]=c^{Q,2}_{mnpq}\dlo{\mu}(\bar{Q}_m\tau^I \gamma_\nu Q_n)\dup{\mu}(\bar{Q}_p\tau^I \gamma^\nu Q_q)$&\\[1mm]
&&$\mathcal{O}_3[Q]=c^{Q,3}_{mnpq}\dlo{\mu}(\bar{Q}_mT^a \gamma_\nu Q_n)\dup{\mu}(\bar{Q}_pT^a \gamma^\nu Q_q)$&\\[1mm]
&&$\mathcal{O}_4[Q]=c^{Q,4}_{mnpq}\dlo{\mu}(\bar{Q}_mT^a\tau^I \gamma_\nu Q_n)\dup{\mu}(\bar{Q}_pT^a\tau^I \gamma^\nu  Q_q)$&\\
\cmidrule{2-3}
% 4d
& \multirow{2}{*}{(4-d)} & $\mathcal{O}_1[d]=c^{d,1}_{mnpq}\dlo{\mu}(\bar{d}_m\gamma_\nu d_n)\dup{\mu}(\bar{d}_p\gamma^\nu d_q)$&\\[1mm]
&&$\mathcal{O}_3[d]=c^{d,3}_{mnpq}\dlo{\mu}(\bar{d}_mT^a\gamma_\nu d_n)\dup{\mu}(\bar{d}_pT^a\gamma^\nu d_q)$&\\
\midrule
\multirow{6}{*}{\vtext{cross-quartic\quad \quad \quad}} & 
% 2u2Q
\multirow{2}{*}{(2-u)(2-Q)} & $\mathcal{O}_{K1}[u,Q]=-a^{uQ,1}_{mnpq}\left(\bar{u}_m\gamma_\mu\dlo{\nu}u_q\right)\left(\bar{Q}_n\gamma^\nu\dup{\mu}Q_p\right)$&\multirow{6}{*}{\vtext{$a^{\psi\chi}_{mnpq}=a^{\chi\psi}_{nmqp}$\qquad\qquad}\quad\vtext{$a_{mnpq}=a^*_{qpnm}$\qquad\qquad}}\\[1mm]
&&$\mathcal{O}_{K3}[u,Q]=-a^{uQ,3}_{mnpq}\left(\bar{u}_mT^a\gamma_\mu\dlo{\nu}u_q\right)\left(\bar{Q}_nT^a\gamma^\nu\dup{\mu}Q_p\right)$&\\
\cmidrule{2-3}
%2d2Q
& \multirow{2}{*}{(2-d)(2-Q)} & $\mathcal{O}_{K1}[d,Q]=-a^{dQ,1}_{mnpq}\left(\bar{d}_m\gamma_\mu\dlo{\nu}d_q\right)\left(\bar{Q}_n\gamma^\nu\dup{\mu}Q_p\right)$&\\[1mm]
&&$\mathcal{O}_{K3}[d,Q]=-a^{dQ,3}_{mnpq}\left(\bar{d}_mT^a\gamma_\mu\dlo{\nu}d_q\right)\left(\bar{Q}_nT^a\gamma^\nu\dup{\mu}Q_p\right)$&\\
%2d2Q
\cmidrule{2-3}
& \multirow{2}{*}{(2-d)(2-u)} & $\mathcal{O}_{K1}[d,u]=-a^{du,1}_{mnpq}\left(\bar{d}_m\gamma_\mu\dlo{\nu}d_q\right)\left(\bar{u}_n\gamma^\nu\dup{\mu}u_p\right)$&\\[1mm]
&&$\mathcal{O}_{K3}[d,u]=-a^{du,3}_{mnpq}\left(\bar{d}_mT^a\gamma_\mu\dlo{\nu}d_q\right)\left(\bar{u}_nT^a\gamma^\nu\dup{\mu}u_p\right)$&\\

\bottomrule
\end{tabular}
\caption{List of independent self-quartic and cross-quartic operators. $T^a$ are the $SU(3)_C$ QCD generators and $\tau^I=\frac{\sigma^I}{2}$ are the $SU(2)_L$ EW generators. }
\label{tab:operators}
\end{center}
\end{table}

Therefore, restricting to the operators listed in Table~\ref{tab:operators}, we can see that there are $2+4+2=8$ independent self-quartic tensors $c_{mnpq}$. As stated already in Ref.~\cite{flavorconstr}, imposing the symmetry requirements $c_{mnpq}=c_{pqmn}$ and $c_{mnpq}=c^*_{nmqp}$ leaves $\frac{1}{2}N_f^2(N_f^2+1)$ independent real entries in each tensor. Indeed, the first condition is a symmetry condition on the complex $N_f^2\times N_f^2$ matrix $c_{mnpq}$ whose rows are indexed by $(m,n)$ and columns by $(p,q)$, so that it leaves $2\times\frac{1}{2}N_f^2(N_f^2+1)$ unconstrained real entries. The second condition further halves them. 
On the other hand, there are $2+2+2=6$ independent cross-quartic structure of operators. Each $a_{mnpq}$ tensor has only to obey the hermiticity condition $a_{mnpq}=a^*_{qpnm}$, thus each of them contains $N_f^4$ independent real entries. Since $a^{\psi\chi}_{mnpq}=a^{\chi\psi}_{nmqp}$, fixing one $a^{\psi\chi}_{mnpq}$ tensor automatically fixes the one with $\psi\leftrightarrow \chi$. Overall, we will deal with $6+8=14$ independent types of operators, and $2N_f^2(5N_f^2+2)$ independent operators overall. 
	
%%%%%%%%%%%%%%%%%%%%%%%%%%%%%%%%%%%%%%%%%%%%%%%%%%%%%%%%%%%%%%%%%%%%%%%%%%%%%%%%%%%%%%%	
	\subsection{MFV ansatz for dimension 8 operators}\label{sec:MFVansatzSection}
	
	We now wish to enforce the MFV assumption on the list of four-fermion dimension-8 operators. This means that all EFT coefficients in Table~\ref{tab:operators} must be written in terms of Yukawa spurions and flavor-blind EFT coefficients.
		
	Since our goal is to study possible bounds on the entries of the Yukawa matrices, then to be compared with phenomenological values, one must in principle depart from the latter and treat the fermion masses and the entries of the CKM matrix as generic. However, this would prevent us from performing a proper power counting, as all powers of those matrices could have in principle the same magnitude. In contrast, phenomenological studies of MFV~\cite{MFV, MFVreview} rely on the measured values of the fermion masses or of the CKM elements to define a consistent expansion. Large Yukawas demand further care, but can also be treated consistently~\cite{Feldmann:2008ja,Kagan:2009bn}. 
We follow the same approach, giving up on full generality and on constraints on the smallest Yukawas; we stick to cases where there exists a Yukawa much larger than the others, so that we can fix all remaining ones to zero at first order in a consistent MFV expansion. This assumption is realized in particular by the phenomenological values of the Yukawas. We will consider two simplified scenarios, with respectively 2 and 3 flavors. Then, the largest Yukawa, $y_c$ in the former case and $y_t$ in the latter, is the only one non-vanishing at leading order. We keep them as free parameters in all the expressions below - see however the discussion at the end of this section for the case of $y_t$. In keeping only the first relevant order in this expansion, we will see that, at least in the proper realizations of $N_f=2,~3$, there is always a choice of basis in flavor space such that the CKM matrix $V_\textit{CKM}$ makes no appearance in the computations, and no hope of putting any bound on its entries can be retained. This is due to the fact that only the up-Yukawa matrix $Y_u$ will enter our expressions, and we can always pick a basis where $V_\textit{CKM}$ is placed exclusively in $Y_d$, the down-Yukawa matrix. This basis is nothing but the one of Eq.~\eqref{eq:interactionbasis}, since in our approximations and when restricted to quark-type Yukawas, it becomes, for $N_f=3$,
\be
	Y_u=\text{diag}(0,0,y_t), \quad Y_d=\text{diag}(0,0,0).
	\label{eq:interactionbasisLeadingOrder}
\ee

The expansion that we use depends on the size of the largest Yukawa. By assumption, we neglect any term where $Y_d$ appears, but an expansion in the up-Yukawa matrix $Y_u$ demands that the entries of the matrix are $<1$, to ensure a consistent, non-divergent expansion. While this works for $N_f=2$ due to the smallness of the charm quark Yukawa, this does not hold for the top, so that the expansion has to be resummed when $N_f=3$. We start by discussing the naive expansion, and discuss at the end of this section how it should be modified to account for the top-Yukawa resummation.

Numerically, for $N_f=3$, and since the Yukawa matrices will always appear in pairs, these approximations amount to neglecting terms of order $\order{\left({y_c}/{y_t}\right)^2}\sim\order{\left({y_b}/{y_t}\right)^2}\sim\order{10^{-3}}$  at most, when setting the Yukawas to their real values. This gives a measure of how much we can let $y_t$ vary without spoiling our approximation. 
In addition, focusing only on quarks is justified, too, at this level. Indeed, since $\left({y_\tau}/{y_t}\right)^2\sim10^{-4}$, the only bilinears formed by leptons that would be added to this order are of the form $\bar{L}_m\Gamma L_m$ or $\bar{e}_m\Gamma e_m$. Thus, they only contribute trivially to the flavor tensor structure, and bounds for the operators built with them can be retrieved, e.g., looking at the ones built with $d$ quark fields. All this is somehow weaker for $N_f=2$. There, the biggest contributions we neglected have an approximate size of $\left({y_s}/{y_c}\right)^2\sim\left({y_\mu}/{y_c}\right)^2\sim10^{-2}$.

	Let us now ask what the MFV ansatz implies for operators containing 4 right-handed up-type quarks, when we work at order $\order{Y_u^2Y_d^0}$ in the expansion. There are two possible operators containing 4 up-quark fields:
	\begin{align*}
	\mathcal{O}_1[u]&=c^{u,1}_{mnpq}\dlo{\mu}(\bar{u}_m\gamma_\nu u_n)\dup{\mu}(\bar{u}_p\gamma^\nu u_q)\\
	\mathcal{O}_3[u]&=c^{u,3}_{mnpq}\dlo{\mu}(\bar{u}_mT^a\gamma_\nu u_n)\dup{\mu}(\bar{u}_pT^a\gamma^\nu u_q),
	\end{align*}
	where only flavor indices are shown. To obtain the MFV expansion of the $c^{u}$ coefficients, it is useful to define two objects\footnote{Obviously, after the spurions freeze to their expectation values, $M=\tilde{ M}$ in our choice of basis.}:
	\begin{align}
	{M}\equiv Y_uY_u^\dagger\ , \\
	\tilde{M}\equiv Y_u^\dagger Y_u\ .
	\label{eq:mmtilde}
	\end{align}
	Let us study what happens for the physical case $N_f=3$ (requiring instead $N_f=2$ can only impose additional constraints that can always be enforced at a later moment). 
	The product of quark bilinears $\sim \bar{u}_m u_n\bar{u}_p u_q$ is a $(\mathbf{\bar{3}}\otimes\mathbf{3})\otimes(\mathbf{\bar{3}}\otimes\mathbf{3})$ of $SU(3)_u$, and can be decomposed as $\mathbf{1}_1\oplus\mathbf{1}_2\oplus\mathbf{8}_1\oplus\mathbf{8}_2\oplus\mathbf{27}$, since the $\mathbf{10}$ and the $\mathbf{\overline{10}}$ vanish because of the exchange symmetry. Then, at $\order{Y_u^0Y_d^0}$, $c_{mnpq}^{u,i}=\rho^{u,i}_1\delta_{mn}\delta_{pq}+\rho^{u,i}_3\delta_{mq}\delta_{pn}$.  Because of the $SU(3)_Q$ index carried by $Y_u$, there is no invariant we can build with just one copy of it. However, the contraction $\tilde{M}$ defined earlier is a singlet of $SU(3)_Q$ and contains a $\mathbf{1}\oplus\mathbf{8}$ of $SU(3)_u$. Its trace can be reabsorbed through a redefinition of the $\rho^{u,i}_1$ and $\rho^{u,i}_3$ coefficients, while with its traceless part we can build two further structures, so that:
	\begin{align}
	c^{u,i}_{mnpq}&=\rho^{u,i}_1(\delta_{mn}\delta_{pq})+\rho^{u,i}_2(\tilde{ M}_{mn}\delta_{pq}+\delta_{mn}\tilde{ M}_{pq})+\rho^{u,i}_3(\delta_{mq}\delta_{pn})+\nonumber\\
	&+\rho^{u,i}_4(\tilde{ M}_{mq}\delta_{pn}+\delta_{mq}\tilde{ M}_{pn}),
	\end{align}
	where all of the $\rho$ coefficients are unconstrained, and can be taken of $\order{1}$.
	In Table~\ref{tab:MFV}, we list the shape that the MFV ansatz forces on the Wilson coefficients respectively of the self-quartic and cross-quartic kinds of operators previously listed. These can be obtained in a similar manner as what was just shown. 

\begin{table}[ht]
\begin{center}
\scalebox{.78}{
\begin{tabular}{ccl}
\toprule
Type & Content & \multicolumn{1}{c}{Operator} \\
\midrule
\multirow{6}{*}{\vtext{self-quartic\quad \quad \quad}} & 
% 4u
(4-u) & \multirow{2}{*}{$c^{u,i}_{mnpq}=\rho^{u,i}_1(\delta_{mn}\delta_{pq})+\rho^{u,i}_2( \tilde{M}_{mn}\delta_{pq}+\delta_{mn} \tilde{M}_{pq})+\rho^{u,i}_3(\delta_{mq}\delta_{pn})+\rho^{u,i}_4( \tilde{M}_{mq}\delta_{pn}+\delta_{mq} \tilde{M}_{pn})$}%\label{eq:uptensor}
\\
& ${}^{i=1,3}$\\
\cmidrule{2-3}
%4Q
& (4-Q) &  \multirow{2}{*}{$c^{Q,i}_{mnpq}=\rho^{Q,i}_1(\delta_{mn}\delta_{pq})+\rho^{Q,i}_2( M_{mn}\delta_{pq}+\delta_{mn} M_{pq})+\rho^{Q,i}_3(\delta_{mq}\delta_{pn})+\rho^{Q,i}_4( M_{mq}\delta_{pn}+\delta_{mq} M_{pn})$}%\label{eq:Qtensor}
\\ 
& ${}^{i=1,2, 3, 4}$\\
\cmidrule{2-3}
% 4d
& (4-d) & \multirow{2}{*}{$c^{d,i}_{mnpq}=\rho^{d,i}_1(\delta_{mn}\delta_{pq})+\rho^{d,i}_3(\delta_{mq}\delta_{pn})$}%\label{eq:downtensor}
\\
& ${}^{i=1, 3}$\\ 
\midrule
\multirow{6}{*}{\vtext{cross-quartic\quad \quad \quad}} & 
% 2u2Q
(2-u)(2-Q) & \multirow{2}{*}{$a^{uQ,i}_{mnpq}=\rho^{uQ,i}_1(\delta_{mq}\delta_{np})+\rho^{uQ,i}_2(\tilde{ M}_{mq}\delta_{np})+\rho^{uQ,i}_3(\delta_{mq} M_{np})+\rho^{uQ,i}_4((Y_u)_{nq}(Y^\dagger_u)_{mp})$}%\label{eq:2u2Qtensor}
\\
& ${}^{i=1, 3}$\\ 
\cmidrule{2-3}
%2d2Q
& (2-d)(2-Q) & \multirow{2}{*}{$a^{dQ,i}_{mnpq}=\rho^{dQ,i}_1(\delta_{mq}\delta_{np})+\rho^{dQ,i}_2(\delta_{mq} M_{np})$}%\label{eq:2d2Qtensor}
\\
& ${}^{i=1, 3}$\\ 
%2d2Q
\cmidrule{2-3}
& (2-d)(2-u) & \multirow{2}{*}{$a^{du,i}_{mnpq}=\rho^{du,i}_1(\delta_{mq}\delta_{np})+\rho^{du,i}_2(\delta_{mq}\tilde{ M}_{np})$}%\label{eq:2d2utensor}
\\
& ${}^{i=1, 3}$\\ 
\bottomrule
\end{tabular}
}
\caption{$\mathcal{O}(Y_u^2Y_d^0)$-MVF expansion of the self-quartic and cross-quartic operators}
\label{tab:MFV}
\end{center}
\end{table}

The above tables define the EFT coefficients on which we will soon apply positivity bounds. The coefficients can be seen to respect all the symmetry properties required by Table~\ref{tab:operators}. However, the number of independent coefficients is drastically reduced, as now the only free parameters are flavor-blind overall coefficients, i.e., the objects we named $\rho_A^{i}$, and their number is independent on $N_f$. In particular, as already stated, the number of independent real coefficients in a unconstrained setting is $2N_f^2(5N_f^2+2)$ (176 for $N_f=2$ and 846 for $N_f=3$), while after imposing MFV we are left with 44 independent real coefficients $\rho_A^{i}$ at the $\mathcal{O}(Y_u^2Y_d^0)$ order in the MFV expansion, independently on the number of flavor (actually, for $N_f=2$, it turns that 6 coefficients are redundant). 

We should nevertheless pause to comment on the $N_f=3$ case. We kept $y_t$ as a generic parameter in our analysis, but are eventually interested in its phenomenological value, which is $\sim\order{1}$. Therefore, the truncation to $\order{Y_u^2Y_d^0}$ we perform is not in principle justified for the physical $N_f=3$ case: higher-order terms such as
	\begin{align}
	\left(Y_uY^\dagger_u\right)^n_{ij}\sim y_t^{2n}  \delta_{3i}\delta_{3j} \ .
	\end{align}
should be properly resummed. Interestingly, the resummation does not bring new flavor violation beyond the one contained in the first non-trivial contraction $Y_uY_u^\dagger$. This is more transparent in the basis of~\eqref{eq:interactionbasis2}, where
	\begin{align}
	\left(Y_uY^\dagger_u\right)^n_{ij}\sim y_t^{2n}  \left(V_\textit{CKM}^*\right)_{3i}\left(V_\textit{CKM}\right)_{3j} \ . 
	\end{align}
This means that the flavor violation structure can be obtained from our \textit{naive} expansion in Table~\ref{tab:MFV}, up to a redefinition of the parameters to account for the resummation. Namely, that means that one should remove the explicit $y_t$ dependence by fixing $y_t=1$ and turn the EFT coefficients into $\order{1}$ arbitrary functions of $y_t$, $\rho_i^{x,j}\rightarrow \rho(y_t)_i^{x,j}$. A proof of such behaviour is given in Appendix~\ref{sec:grouptheoryLargeTop}. When we freeze the spurions to their background values, functions of $y_t$ become simple numbers, which means that any explicit $y_t$-dependence simply disappears from the expansion.

%%%%%%%%%%%%%%%%%%%%%%%%%%%%%%%%%%%%%%%%%%%%%%%%%%%%%%%%%%%%%%%%%%%%%%%%%%%%%%%%%%%%%%%	
\section{Analysis of the bounds}
	\label{sec:studyingthebounds}
	At this point, we have all the machinery we need to pursue our goal, namely to confront  the bounds obtained in Ref.~\cite{flavorconstr} with the MFV hypothesis dictating the expansion of the various dimension-8 operators in power of the Yukawa, as listed in Table~\ref{tab:MFV}. First of all, the positivity constraints of Ref.~\cite{flavorconstr} depend not only on the Wilson coefficients $c_{mnpq}$ and $a_{mnpq}$, but also on some arbitrary external states, dubbed $\alpha$ and $\beta$ and consisting in generic complex vectors of unit norm in flavor space. Thus, we need to disentangle the former from the latter first, to obtain expressions that depend on the operator coefficients only. 
	Secondly, many of the inequalities contain linear combinations of coefficients coming from distinct independent operators. To simplify the computations, we define new coefficients via suitable linear transformations.  
	We will carry this latter simplification first, and then proceed to show how we removed the dependencies on the $\alpha$'s and $\beta$'s.
	
	Let us see, for example, how this works in the case of operators containing 4 up fields. 
	The bounds on them are obtained~\cite{flavorconstr} by scattering the following states:
	\begin{align}
		\ket{\psi_1}&=\alpha_{mi}\ket{\bar{u}_{mi}}, &\ket{\psi_2}&=\beta_{mi}\ket{u_{mi}}, \nonumber\\
		\ket{\psi_3}&=\beta^*_{mi}\ket{\bar{u}_{mi}}, &\ket{\psi_4}&=\alpha^*_{mi}\ket{u_{mi}},
	\end{align}
	where $m$ and $i$ are flavor and gauge indices respectively.
	The amplitude then reads:
	\begin{align}
	\mathcal{A}=4s^2\left[\left(c^{u,1}_{mnpq}-\frac{1}{6}c^{u,3}_{mnpq}\right)\alpha_{mi}^*\beta_{ni}\beta^*_{pj}\alpha_{qj}+\frac{1}{2}c^{u,3}_{mnpq}\alpha_{mi}^*\beta_{nj}\beta^*_{pj}\alpha_{qi}\right].
	\end{align}
	Marginalizing over the gauge indices\footnote{We factor out from now on the gauge dependence from the $\alpha$'s and $\beta$'s, so that they only contain flavor indices. This yields conservative bounds, albeit not necessarily the strongest ones.}, two different bounds are obtained:
	\begin{align}
	\alpha_m\alpha^*_q\beta_n\beta^*_p \left(c_{mnpq}^{u,1}+\frac{1}{3}c_{mnpq}^{u,3}\right)&>0\, ,\nonumber\\
	\alpha_m\alpha^*_q\beta_n\beta^*_p c_{mnpq}^{u,3}&>0 \ .
	\label{eq:4uboundsfirst}
	\end{align}
	Here, as mentioned, $\alpha_n$ and $\beta_n$ are arbitrary $N_f$ components complex vectors of unit norm. They parametrize the external states, since the bounds are obtained by constraining the $s^2$ coefficient of a $2\to2$ scattering of generic superpositions of flavor eigenstates. Therefore, the inequalities in Eq.~\eqref{eq:4uboundsfirst} have to be fulfilled for all 
$\alpha$'s, $\beta$'s.
		
We perform a linear transformation on Eq.~\eqref{eq:4uboundsfirst} by defining:
	\begin{align}
	\xi^{u,1}_k \equiv\rho^{u,1}_k+\frac{1}{3}\rho^{u,3}_k \quad \textrm{and} \quad \xi^{u,3}_k\equiv\rho^{u,3}_k \quad \textrm{for}\quad k=1,2,3,4,
	\label{eq:upredefinition}
	\end{align}
so that, defining $c(\xi)^{u,i}_{mnpq}$ as in the first line of Table~\ref{tab:MFV} but with $\rho\to\xi$, i.e., 
	\begin{align*}
	c(\xi)^{u,i}_{mnpq}&=\xi^{u,i}_1(\delta_{mn}\delta_{pq})+\xi^{u,i}_2(\tilde{ M}_{mn}\delta_{pq}+\delta_{mn}\tilde{ M}_{pq})+\xi^{u,i}_3(\delta_{mq}\delta_{pn})\nonumber\\
	&+\xi^{u,i}_4(\tilde{ M}_{mq}\delta_{pn}+\delta_{mq}\tilde{ M}_{pn}) & i&=1,3,
	\end{align*}
	the bounds become simply 
	\begin{align}
	\alpha_m\alpha^*_q\beta_n\beta^*_p c(\xi)_{mnpq}^{u,i}&>0, & i&={1,3}.
	\label{eq:upbounds}
	\end{align}
	Since all the bounds are expressed as inequalities on linear combinations of the flavor structure tensors as in Eq.~\eqref{eq:4uboundsfirst}, it is always possible to perform a linear redefinition such as Eq.~\eqref{eq:upredefinition} to bring them to a form analogous to Eq.~\eqref{eq:upbounds}. From now on, we will do this  on all operators, and show both bounds and flavor tensors as functions of $\xi$'s. Their explicit dependence on the original $\rho$ coefficients is shown in Appendix~\ref{sec:rhodependence}.
In conclusion, the bounds we have to study are all of the form:
	\begin{IEEEeqnarray}{0rCl"lCl}
	\alpha_m\alpha^*_q\beta_n\beta^*_p c(\xi)_{mnpq}^{X,i}&>&0, & X&=&u, Q, d,
	\label{eq:selfbounds}\IEEEeqnarraynumspace\\
	\alpha_m\alpha^*_q\beta_n\beta^*_p a(\xi)^{X,i}_{mnpq}&>&0, & X&=&uQ, dQ, du.
	\IEEEeqnarraynumspace
	\label{eq:crossbounds}
	\end{IEEEeqnarray}	
We first prove that it is always possible to find some $\xi$'s such that these constraints can be satisfied for any $\alpha$, $\beta$. Indeed, for the self-quartic operators, one can notice that, looking at the pattern in which the flavor indices are summed, if we choose $\xi_1^i=\xi_2^i=0$, then the bounds can be expressed as
	\begin{align}
	\xi_3^i  |\alpha|^2|\beta|^2 + \xi_4^i\left(\alpha_m A_{mq}\alpha_q^*|\beta|^2 + \beta^*_p A_{pn}\beta_n|\alpha|^2\right) >0,
	\end{align}
	where $A=\tilde{ M}$, $A= M$ and $A=0$ for the (4-u), (4-Q) and (4-d) cases respectively. 
	In the former two cases, being the product of an invertible matrix and its hermitian conjugate, $A$ is (semi-)positive definite. Thus $\xi_1^i=\xi_2^i=0$, $\xi_3^i>0$ and $\xi_4^i>0$ is, in this setting, an allowed region in the parameter space, and fulfills the bounds $\forall \alpha,\beta $. In the (4-d) case, $A=0$, and $\{\xi_1^i=0,\xi_3^i>0\}$ is an always allowed region of the parameter space. Similar conclusions can be drawn for the cross-quartic operators. As a consequence, there exists at least one region that is a solution of Eqs.~\eqref{eq:selfbounds} and~\eqref{eq:crossbounds}, with the coefficients expressed as per Table~\ref{tab:MFV}. 
	
However, we wish to study the anatomy of the bounds when all the $\rho$ coefficients of the MVF expansions are of $\mathcal{O}(1)$, which is the natural realization of the MFV ansatz.  By fixing the coefficients, the bounds become functions of SM parameters alone. We then verify whether they are strict enough as to impose constraints on the parameters of the dimension-4 Lagrangian. In the two flavor case, that means $y_c$, the charm-quark Yukawa coupling. Naively, that also means $y_t$ when $N_f=3$, however, as we discussed previously, $y_t$ should be absorbed in the $\rho$  (equivalently, in the $\xi$) coefficients. In that case, we can only check whether all the EFT coefficients can be consistently $\mathcal{O}(1)$.
	
%%%%%%%%%%%%%%%%%%%%%%%%%%%%%%%%%%%%%%%%%%%%%%%%%%%%%%%%%%%%%%%%%%%%%%%%%%%%%%%%%%%%%%%	
\subsection{Flavor violation and CKM-(in)dependence of the positivity bounds}
\label{sec:flavorviolation}

	Before going any further, some clarifications are in order. All along the discussion we made, it looks like there is no place for any flavor violation at all. Indeed, in our approximation, the only matrix involved in building the flavor invariants is $Y_u$, which we chose to be diagonal, and the only physical parameter shaping the allowed region is $y_t$, while there is no sign of the CKM matrix.
	Obviously, our discussion cannot depend on the specific basis that we pick. In this section we show that this is the case and that the CKM matrix only enters the bounds at subleading order with respect to our approximations. Imagine we chose, instead of Eq.~\eqref{eq:interactionbasis}, the basis~\eqref{eq:interactionbasis2}. Then, we could have diagonalized $Y_u$, and consequently $M$ and $\tilde{ M}$, at a later time, by exploiting the redundancy contained in the definition of expressions like Eq.~\eqref{eq:selfbounds} or~\eqref{eq:crossbounds}. Indeed, we know that any square matrix can be decomposed as
	\begin{align}
	Y_u=U \Sigma \tilde{U}^\dagger,
	\label{eq:ydiagonaliz}
	\end{align}
	where $U$ and $\tilde{U}$ are unitary matrices and $\Sigma$ is diagonal. Then:
	\begin{align}
	M&=Y_u Y_u^\dagger=U \Sigma \Sigma^* U^\dagger\\
	\tilde{ M}&=Y_u^\dagger Y_u =\tilde{U} \Sigma^*\Sigma  \tilde{U}^\dagger.\\
	\end{align}
Thus, $M$ and $\tilde{ M}$ are diagonalized by $U$ and $\tilde{U}$ respectively.
Then, in Eq.~\eqref{eq:selfbounds} or~\eqref{eq:crossbounds}, we could have rotated\footnote{More precisely, we can multiply by the identity $\mathbbm{1}_{N_f}=U U^\dagger $ so that \begin{align*}
	&\alpha_m\alpha^*_q\beta_n\beta^*_p c_{mnpq}^{u,i}=\\
	&=\alpha_{m}\left(U_{mm'}U^\dagger_{m'm''}\right)\left(U_{q''q'}U^\dagger_{q'q}\right)\alpha^*_{q}\left(U_{n''n'}U^\dagger_{n'n}\right)\beta_n\beta^*_p\left(U_{pp'}U^\dagger_{p'p''}\right)c_{m''n''p''q''}\equiv\\
	&\equiv\left(\tilde{\alpha}_{m'}U^\dagger_{m'm''}\right)\left(U_{q''q'}\tilde{\alpha}^*_{q'}\right)\left(U_{n''n'}\tilde{\beta}_{n'}\right)\left(\tilde{\beta}^*_{p'}U^\dagger_{p'p''}\right)c_{m''n''p''q''}
	\end{align*}} both $\alpha$ and $\beta$ (and their hermitian conjugates) using $U$ or $\tilde{U}$. This does not modify the space that $\alpha$ and $\beta$ span, since unitary matrices conserve norms. Therefore we can explore the  $\alpha$'s and $\beta$'s space with the diagonalized version of $M$ and $\tilde{ M}$. In particular, $\tilde{ M}=\lambda_u^2$ is diagonal to begin with, while we can rotate $M\to\Sigma^*\Sigma= \lambda_u^2$ using $U=V_\textit{CKM}^\dagger$,  $\tilde{U}=\mathbbm{1}_{N_f}$. These are the same matrices we got when we started with the basis in Eq.~\eqref{eq:interactionbasis} in the first place. The freedom to absorb unitary matrices in the generic vectors $\alpha$ and $\beta$ arises from specific properties of the positivity bounds. First, those bounds are obtained in Ref.~\cite{flavorconstr} in a high-energy limit where all SM fermions are considered massless. In this limit, the mass terms disappear and they do not single out anymore the preferred basis that diagonalizes them. In addition, only dimension-8 operators are constrained by the bounds, so that the dimension-8 EFT coefficients are the only spurions that break the flavor symmetry and enter the bound. Thus, the flavor symmetry can be used to absorb \textit{irrelevant} parameters, here in the sense of \textit{not entering the positivity bounds}, among the ones that form the dimension-8 EFT coefficients. In our case, the CKM matrix is precisely such an \textit{irrelevant} parameter. Notice that this statement derives from the use of the full flavor group. Consequently, it does not hold if we only scatter a subset of the flavor states (said differently, if we imposed some conditions on $\alpha,\beta$). Indeed, the restriction of the flavor group to those states may not be sufficient to remove all the CKM dependence from the bounds. We will see an example of this in Section~\ref{sec:nf2revisisted}, when a two-flavor scenario is embedded in $N_f=3$. There, by scattering the two first flavors only, we obtain a subset of the $N_f=3$ bounds which depends on the entries of the CKM matrix. Nevertheless, the full $N_f=3$ bounds are more stringent and do not depend on the latter.
	
	One can then ask at which order in the expansion in powers of Yukawa matrices a CKM contribution would appear so that it could not be rotated away.
	From what we just saw, this has to be a combination containing both up- and down-Yukawa matrices. If we define $N_{mn}\equiv {(Y_dY_d^\dagger)}_{mn}$, i.e., the analogous of $M$ for the down Yukawa matrix, we notice that this, too, contains a $\mathbf{8}$ of $SU(3)_q$. Thus, if we expand a bit further, we can add for example to the second line of Table~\ref{tab:MFV} a term like $\sim \tilde{\rho}_1 (N_{mn}\delta_{pq}+\delta_{mn} N_{pq})+\tilde{\rho}_2(n\leftrightarrow q)$. 
	With this example we can see that the freedom left by the redundancy in the definition of the external states is larger than the symmetry of the Lagrangian alone. Indeed, by exploiting the flavor $U(3)^3$, we can diagonalize either $N$ or $M$, but not both.
	However, from the point of view of the bounds, adding only the aforementioned terms corresponds to a shift $M\to M+N$ in the second line of Table~\ref{tab:MFV}. This matrix can then be diagonalized, since it is hermitian, too. Nonetheless, its eigenvalues depend now explicitly on the entries of the CKM matrix, that consequently enter the bounds in any case at this level in the expansion. 
	
	Conversely, the fact that the entries of $V_\textit{CKM}$ are relevant only at such a subleading order means that the bounds are not really sensitive to their values, and even relatively large modifications for them do not affect much the structure of the bounds.

%%%%%%%%%%%%%%%%%%%%%%%%%%%%%%%%%%%%%%%%%%%%%%%%%%%%%%%%%%%%%%%%%%%%%%%%%%%%%%%%%%%%%%%	
	\subsection{Disentangling the external states.}
	
	As already stated\footnote{We focus here on the $c_{mnpq}$ as a generalization to the $a_{mnpq}$ is straightforward. For the sake of simplicity, and since this analysis applies everywhere, we here drop any superscript on $c_{mnpq}$.}, positivity conditions like Eq.~\eqref{eq:upbounds} have to be fulfilled for every value of $\alpha$ and $\beta$, since they simply label arbitrary in-states. However, to obtain bounds that are purely expression of the EFT coefficients, one has to disentangle the latter from the external states. This section is devoted to show how this can be done in the case under consideration. We can start by removing the dependence of the bounds on either $\alpha$ or $\beta$. 
	Suppose we fix $\beta$ and define $C(\beta)_{mq}=c_{mnpq}\beta_n\beta^*_p$. Notice that this matrix is hermitian, and thus diagonalizable, and even if it were not, its antihermitian part would drop out of expressions like Eq.~\eqref{eq:upbounds}. Then the positivity requirement~\eqref{eq:upbounds} takes the form $C(\beta)_{mq}\alpha_m\alpha^*_q>0$, and this inequality has to be satisfied for any unit $N_f$-vector $\alpha$. This is equivalent to asking that the  matrix $C(\beta)$ is positive definite, i.e.,  that its real eigenvalues $r(\beta)_{I}$, $I=1,\ldots, N_f$, are all positive.
	Thus, we can trade 
	\begin{align}
	\begin{cases}
	c_{mnpq}\beta_n\beta^*_p\alpha_m\alpha^*_q>0\\
	\forall \alpha, \beta\  \textrm{with } \norm{\alpha}=\norm{\beta}=1
	\end{cases}
	\iff
	\begin{cases}
	r(\beta)_{I}>0\quad I=1,\ldots, N_f\\
	\forall \beta\  \textrm{with }  \norm{\beta}=1
	\end{cases}
	\ .
	\label{eq:equivalentconditions}
	\end{align}

	The conditions on the r.h.s of Eq.~\eqref{eq:equivalentconditions} are necessary and sufficient. They are necessary since, if we find a negative eigenvalue for some $\beta=\hat{\beta}$, we can pick $\alpha=\hat{\alpha}$ to be an eigenvector associated to that eigenvalue and the quartic expression on the l.h.s of Eq.~\eqref{eq:equivalentconditions} evaluated at $\hat{\alpha},~\hat{\beta}$ would be negative. They are sufficient because, if there is some value of $\alpha$, and $\beta$, say $\hat{\alpha},~\hat{\beta}$, in which the l.h.s of Eq.~\eqref{eq:equivalentconditions} is negative, or in other words $C(\hat{\beta})_{mq}\hat{\alpha}_m\hat{\alpha}^*_q<0$, then $C(\hat{\beta})$ has to have at least one negative eigenvalue. Indeed, one can decompose $\hat{\alpha}$ on the basis $\{v^I_m\}$ of eigenvectors of $C(\hat{\beta})$, i.e., write $\hat{\alpha}_m=\hat{\alpha}_I v^I_m$, and obtain $C(\hat{\beta})_{mq}\hat{\alpha}_m\hat{\alpha}^*_q=\sum_I r(\beta)_{I}\abs{\hat{\alpha}_I}^2\norm{v^I}^2<0$.
		
Another way to phrase the conditions in the r.h.s of Eq.~\eqref{eq:equivalentconditions} is by noticing that expressions like Eq.~\eqref{eq:upbounds} can be viewed as quadratic homogeneous polynomials in the complex components of $\alpha$, with zero linear term. Then, requiring that the polynomial is greater than zero reduces to asking the multidimensional parabola to point upwards in any direction parametrized by $\alpha$. Were this not the case, we could find an eigendirection with negative hessian eigenvalue, along which we would end up in the negative region.

	Since $\alpha$ and $\alpha^*$ always appear in pairs, and so do $\beta$ and $\beta^*$, we can remove a total phase from each of them. Moreover, they are of fixed unit norm. Then, they contain $2N_f-2$ free real parameters each, and the l.h.s of Eq.~\eqref{eq:equivalentconditions} depends on $4N_f-4$ parameters. We trade it for the $N_f$ conditions on the eigenvalues, each condition depending only on the $2N_f-2$ real parameters contained in $\beta$. This rapidly turns out to be inconvenient for large values of $N_f$, but it works well for $N_f=2,~3$.
	In particular, for $N_f=2$, the two eigenvalues are positive if and only if the trace and the determinant of $C(\beta)_{mq}$ are positive. Notice that the discrepancy in the counting between the two requirements lies only in the number of free parameters we have to marginalize over. Instead, having shown that the r.h.s. and the l.h.s. of Eq.~\eqref{eq:equivalentconditions} are equivalent, they impose the same conditions on the $c_{mnpq}$ after all the $\alpha$'s and $\beta$'s are removed.
	
%%%%%%%%%%%%%%%%%%%%%%%%%%%%%%%%%%%%%%%%%%%%%%%%%%%%%%%%%%%%%%%%%%%%%%%%%%%%%%%%%%%%%%%
	\subsection{A benchmark case: (4-Q) operators}
	\label{sec:studyselfquartic}
	We start by studying the positivity bounds~\eqref{eq:selfbounds} for the (4-Q) operators, and work it out step by step, the procedure for the other cases being very similar.  Under the MFV assumption, the coefficients of the self-quartic (4-Q) operators take the form:
	\begin{align*}
	c(\xi)^{Q,i}_{mnpq}&=\xi^{Q,i}_1(\delta_{mn}\delta_{pq})+\xi^{Q,i}_2(M_{mn}\delta_{pq}+\delta_{mn}M_{pq})+\xi^{Q,i}_3(\delta_{mq}\delta_{pn})+\nonumber\\
	&+\xi^{Q,i}_4(M_{mq}\delta_{pn}+\delta_{mq}M_{pn}), & i&={1,2,3,4} .
	\end{align*}
In order to see what are the consequences imposed by Eq.~\eqref{eq:selfbounds} on the coefficients $\xi^{Q,i}_A$,
we will first consider for simplicity  a 2-flavor setting which can be implemented in two slightly different ways: it will successively describe a theory of two generations only (Section~\ref{sec:4upnf2}), and the restriction of a 3-flavor setting to the lightest two flavors (Section~\ref{sec:nf2revisisted}). Then, we proceed to study $N_f=3$.

%%%%%%%%%%%%%%%%%%%%%%%%%%%%%%%%%%%%%%%%%%%%%%%%%%%%%%%%%%%%%%%%%%%%%%%%%%%%%%%%%%%%%%%	
\subsubsection{Positivity bounds on true $N_f=2$ ansatz}
\label{sec:4upnf2}

First of all, if we reduce the symmetry group to be $SU(2)$, one can verify that $(M_{mq}\delta_{pn}+\delta_{mq}M_{pn})$ is not an independent structure, and its coefficient $\rho_4$ can be reabsorbed through a redefinition of the remaining three.
This can be seen both by counting the allowed singlets in the tensor product, or in a more direct way, as shown in Appendix~\ref{sec:grouptheoryrho4}. This way, we can remove 4 $\xi_4^{Q,i}$ coefficients. However, since they provides just an innocuous redundancy, we will keep them at first and set them to zero at a later moment, highlighting the physical consequences of the two choices. Now, we need to parametrize the generic complex unit vector $\beta\in\mathbb{C}^2$. A possible parametrization is:
	\begin{align}
	\beta&=
	\begin{pmatrix}
	x e^{i \theta_x}\\
	y e^{i \theta_y}\\
	\end{pmatrix}\equiv
	e^{i \theta_y}\begin{pmatrix}
	x e^{i \tilde{\theta}_x}\\
	y\\
	\end{pmatrix}& \textrm{with}\quad x^2+y^2&=1,
	\label{eq:betaparn2}
	\end{align}
where all the parameters are real positive and $\tilde{\theta}_x=\theta_x-\theta_y$. As mentioned, we can remove the total phase and set $\theta_y=0$. In the flavor basis~\eqref{eq:interactionbasis}, the up-Yukawa matrix is simply
	\begin{align}
	Y_u&=
	\begin{pmatrix}
	y_u & 0\\
	0 & y_c
	\end{pmatrix} \sim y_c \begin{pmatrix}
	0 & 0\\
	0 & 1
	\end{pmatrix}, 
	\end{align}
$y_u$, $y_c$ being the up and charm Yukawa respectively, and in the last equality, we specifically assumed a mass hierarchy and kept only the leading term. Although we also set $y_c$ to zero in the $N_f=3$ case, we keep it here since it corresponds to the largest mass of this two-flavor theory. Now, as anticipated, we can translate the positivity condition~\eqref{eq:selfbounds} as two conditions on the eigenvalues of $C(\beta)_{mq}=c_{mnpq}\beta_n\beta^*_p$, or, equivalently, on its determinant and trace. 

\begin{figure} 
		\centering
		\textbf{Bounds on (4-Q) operators coefficients for $N_f=2$}\\\medskip
		\includegraphics[width=0.5\textwidth]{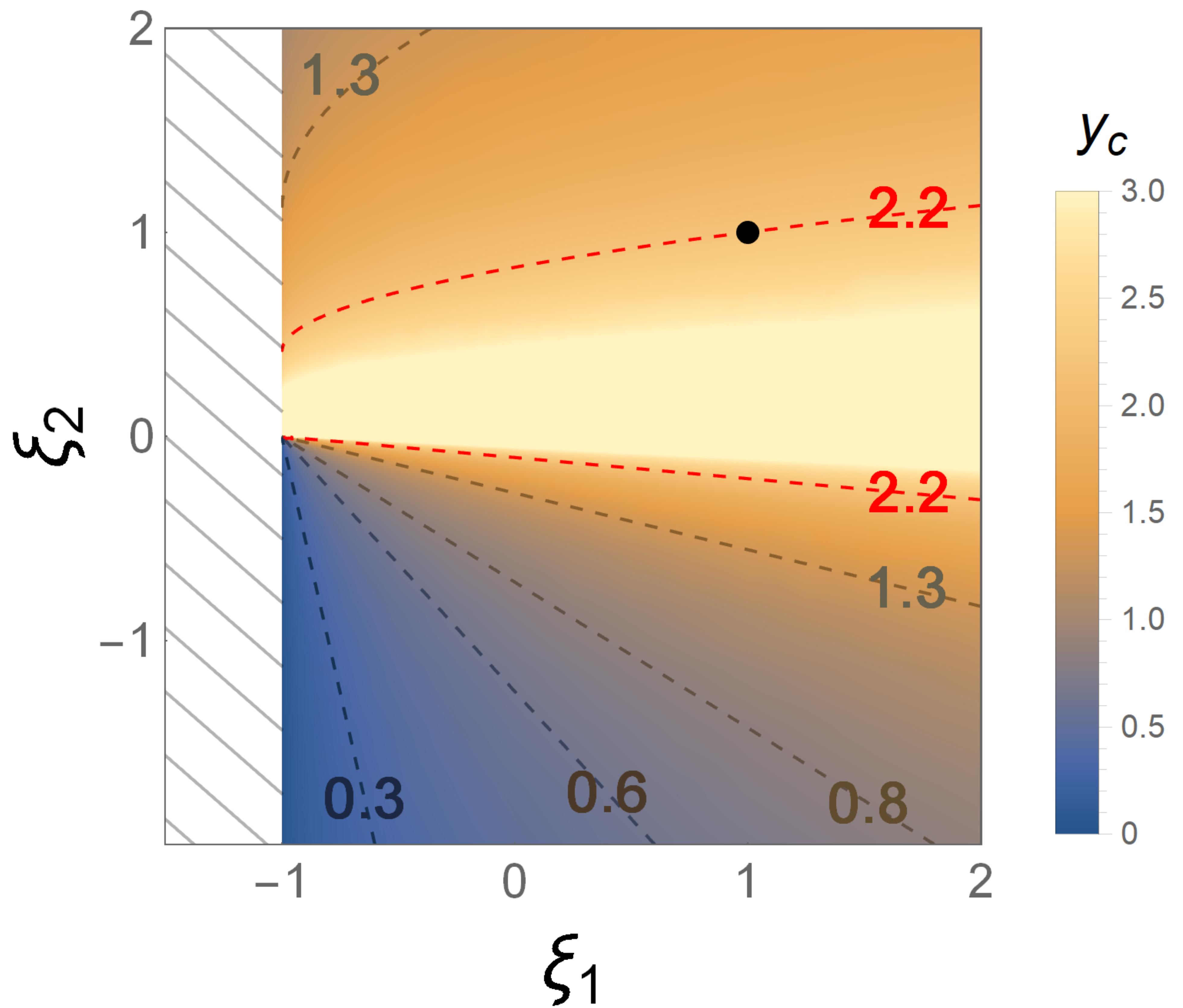}
		\caption{Plot showing the allowed region (in color) obtained for the (4-Q) operators restricted to $N_f=2$ with generic $\xi$ values, as $y_c$ changes. Every region associated to a larger $y_c$ value is contained in the previous ones: for instance, blue and dark orange regions are allowed for any $y_c$ roughly smaller than 4, but forbidden for larger values of $y_c$. The redundant $\xi_4$ has been set to 0. In this case $\xi_3>0$, and using that the bounds are invariant under a full rescaling, we have set $\xi_3=1$, and plot the remaining two independent coefficients. As explained in the text, values of $y_c>1$ are unphysical and are only plotted for visual reasons. The black point represents the \textit{natural} MFV benchmark point $\xi_{1,2,3}=1$. The red line contours the region corresponding to the threshold value of $y_c=\sqrt{2(1+\sqrt{2})}$: for bigger values of $y_c$, the \textit{natural} benchmark point does not belong to the allowed region any more.  The region $\xi_1<-1$ is excluded for any value of $y_c$. For the physical value $y_c\approx 10^{-2}$, almost all points $(\xi_1\geq -1,\xi_2)$ are allowed.}
\label{fig:4QNF2projections}
\end{figure}

The assumed mass hierarchy, $y_c\gg y_u\sim0$, ensures that  the trace and the determinant of $C(\beta)$ depend only on $x$ and not on $y$ nor on $\tilde{\theta}_x$. They are given by:
	\begin{align}
	\Tr[C(\beta)]&=2 x^2 \left(\xi^{Q,i}_2+\xi^{Q,i}_4\right) y_c^2+\xi^{Q,i}_4 y_c^2+\xi^{Q,i}_1+2 \xi^{Q,i}_3\\
	\det[C(\beta)]&=x^4 \left(\xi^{Q,i}_2+\xi^{Q,i}_4\right)^2 y_c^4-x^2 y_c^2 \left(\xi^{Q,i}_2+\xi^{Q,i}_4\right) \left(y_c^2 \left(\xi^{Q,i}_2-\xi^{Q,i}_4\right)-2 \xi^{Q,i}_3\right)+\nonumber\\
	&\quad+\left(\xi^{Q,i}_4 y_c^2+\xi^{Q,i}_3\right) \left(\xi^{Q,i}_1+\xi^{Q,i}_3\right).
	\end{align}
The trace is a linear function of $x^2$ which varies within the interval $[0,1]$. Thus, it is positive for any relevant value of $x$ if and only if its values at the boundaries of the interval are also positive. The determinant, on the other hand, is a quadratic polynomial in $x^2$. One can verify that such parabola is positive in $[0,1]$ if and only if\footnote{This is jusfied like this: assuming it is positive in 0 and 1: if it has negative discriminant, it is positive in the whole interval. Otherwise, if the discriminant is positive, and if the parabola opens downwards, i.e., $a<0$, it is also positive within the interval. If $\Delta>0$ and $a>0$, we then need to make sure that the minimum falls outside $[0,1]$. This is done by requiring $x_{min}^2-x_{min}>0\longrightarrow b(b+2a)>0$.}:
	\begin{itemize}
		\item it is positive at the boundaries;
		\item one of the following conditions is met:
		\begin{align*}
		\Delta<0 \text{ or } a<0 \text{ or } b(b+2a)>0,
		\end{align*}
where we parametrized $\det[C(\beta)]\equiv ax^4+b x^2+c$.
	\end{itemize}
Putting everything together, and after some simplifications, we get the full set of conditions:
	\begin{align}
	\text{(4-Q) ($N_f$=2):  }
	\begin{cases}
	\bullet \ \xi^{Q,i}_4 y_c^2+\xi^{Q,i}_3>0,\\
	\bullet \ 2 y_c^2 \left(\xi^{Q,i}_2+\xi^{Q,i}_4\right)+\xi^{Q,i}_1+\xi^{Q,i}_3>0,\\
	\bullet \ \xi^{Q,i}_1+\xi^{Q,i}_3>0,\\
	\bullet \ y_c^4 (\xi^{Q,i}_4-\xi^{Q,i}_2) \left(\xi^{Q,i}_2+3 \xi^{Q,i}_4\right)+8 \xi^{Q,i}_3 \xi^{Q,i}_4 y_c^2+4 \left(\xi^{Q,i}_3\right)^2>0\quad \textrm{or}\\ %\lor \\
	\quad \left(-4 y_c^2 \left(\xi^{Q,i}_1 \xi^{Q,i}_4+\xi^{Q,i}_2 \xi^{Q,i}_3\right)+y_c^4\left(\xi^{Q,i}_2-\xi^{Q,i}_4\right)^2-4 \xi^{Q,i}_1 \xi^{Q,i}_3\right)<0.
	\end{cases}
	\label{eq:4uN2bounds}
	\end{align}
The allowed region specified by these bounds is shown in Fig.~\ref{fig:4QNF2projections} as a function of the unique relevant parameter $y_c$. 

One can notice in particular that the \textit{natural} MFV benchmark point $\xi_{1,2,3}=1$ is compatible with the positivity bound~\eqref{eq:selfbounds} if and only if $y_c^2<2 \left(1+\sqrt{2}\right)$. However, as mentioned already, a consistent MFV expansion in $Y_u$ requires $y_c<1$. So for any consistent MVF expansion, the positivity bounds are easily satisfied.
	
%%%%%%%%%%%%%%%%%%%%%%%%%%%%%%%%%%%%%%%%%
\subsubsection{$N_f=2$ revisited: projected $N_f=3$ onto $N_f=2$}
\label{sec:nf2revisisted}

Another approach one could follow to describe the $N_f=2$ case, is to take the full $N_f=3$ setting, and to restrict all flavor indices to be $\{1,2\}$. In our flavor basis~\eqref{eq:interactionbasis}, this turns out to be trivial, and it leads to the same result as depicted in Fig.~\ref{fig:4QNF2projections}. However, in the present case, one can modify the bounds using a $U(3)^3$ transformation: although the full $N_f=3$ bounds (to be discussed in the next section) are basis-independent, what we identify with the first two flavors is a basis dependent statement, and so are the bounds derived using the restricted $N_f=2$ approach of this section. As an example, we rotate to the basis of Eq.~\eqref{eq:interactionbasis2}, and only then perform the projection. Consequently, the restricted version of $M$ is now:
	\begin{align*}
	M_{ij}&=\left(Y_uY^\dagger_u\right)_{ij}\sim (V_\textit{CKM})_{3i}(V_\textit{CKM}^*)_{3j} \ , & i,j&=1,2 \ ,
	\end{align*}
	while $\tilde{ M}=0$. In comparison with the usual value of $M_{ij}\sim y_t^2 (V_\textit{CKM})_{3i}(V_\textit{CKM}^*)_{3j}$, we fixed $y_t=1$ to account for the $y_t$-resummation, as explained at the end of Section~\ref{sec:MFVansatzSection}.
	At this point, we are left only with a $U(2)^3$ symmetry, part of which, $U(2)_Q$ for the present (4-Q) case, can be used to diagonalize $M$. This can always be done since $M$ is still hermitian. 
	Here we parametrize $V_\textit{CKM}$ through the Wolfenstein parametrization up to order $\mathcal{O}(\lambda^5)$, where $\lambda=\sin(\theta_c)$, $\theta_c$ being the Cabibbo angle:
	\begin{align*}
	V_\textit{CKM}\approx\left(
	\begin{array}{ccc}
	1-\frac{1}{2} \lambda ^2 -\frac{1}{8}\lambda ^4 &\  \lambda  &\  A \lambda ^3 (\rho -i \eta ) \\
	-\lambda +\frac{1}{2} A^2 \lambda ^5 [1-2(\rho +i\eta)] &\  1 -\frac{1}{2}\lambda ^2 -\frac{1}{8}  \lambda ^4\left(1+4 A^2\right)&\  A \lambda ^2 \\
	A \lambda ^3 [1-(1-\frac{1}{2}\lambda^2)(\rho+i\eta)]&\  -A\lambda^2 +\frac{1}{2}A\lambda^4[1-2(\rho+i\eta)]&\  1-\frac{1}{2}A^2 \lambda ^4 \\
	\end{array}
	\right).
	\end{align*}
	Diagonalizing the 2$\times2$ matrix $M$, one can see that its only non-zero eigenvalue is:
	\be
	\sigma\equiv A^2 \lambda ^4 ,
	\ee
which again we require to be $<1$ to ensure a consistent expansion of the unitary $V_{CKM}$ matrix.
One can then easily map the positivity bounds on the $\xi_{1,2}$ parameter space using the results of the previous section
by substituting $y_c^2$ by $\sigma$. Note that the values of $\xi_{1,2}$ compatible with the positivity bounds now depend on $\sigma$, which itself depends explicitly on the CKM entries, contrary to the general property presented in Section~\ref{sec:flavorviolation}. Indeed, when going back to how the bounds where found in the first place, we see that the setting studied in this section corresponds to a $2\to2$ scattering where the initial and final states are restricted to the first two flavors.  However, fixing them breaks the flavor symmetry down to $U(2)^3$. The latter is then too small to absorb all the CKM parameters, which consistently enter the bounds.  This is different in the full $N_f=3$ case, as we now discuss.

%%%%%%%%%%%%%%%%%%%%%%%%%%%%%%%%%%%%%%%%%
\subsubsection{$N_f=3$}
\label{sec:4upnf3}

Now we wish to tackle the $N_f=3$ setup. We can approximate $M=\left(Y_uY^\dagger_u\right)_{ij}\sim \delta_{3i}\delta_{3j}$, $i,j=1,2,3$, again after fixing $y_t=1$. The only non-zero eigenvalue of this matrix is obviously $1$.
Barring a total irrelevant phase, we can parametrize the complex unit vector $\beta\in \mathbb{C}^3$ as
	\begin{align}
	\beta&=
	\begin{pmatrix}
	x e^{i \theta_x}\\
	y e^{i \theta_y}\\
	z \\
	\end{pmatrix},\quad \textrm{with}\quad x^2+y^2+z^2=1.
	\end{align}
As before, the positivity bounds~\eqref{eq:selfbounds} mapped onto the $\xi^{Q,i}_{1,2,3,4}$  space will be obtained by requiring that the eigenvalues of the matrix $C(\beta)$ are all positive. For simplicity, let us first compute these eigenvalues for the \textit{natural} benchmark point with all $\xi=1$. Again, because of the mass hierarchy, $y_t \gg y_c, y_u$, the characteristic polynomial depends only on $z$ and not on $x,y, \theta_x, \theta_y$. It factorizes nicely:
	\begin{align}
	p(t)=&-\left(t-z^2-1\right) \left[t^2-4t \left(z^2+1\right)+4\left(1+ z^2+z^4\right)\right],
	\end{align}
so its first eigenvalue is simply $t_1(z)= z^2+1$ and it is always positive.
To avoid unpleasant radicals, we can evaluate the sum and product of the remaining two eigenvalues. This is equivalent to taking the trace and the determinant of $C(\beta)$ and subtract and factor out $t_1$ respectively:
	\begin{eqnarray}
	t_2(z)+t_3(z)=4\left(z^2+1\right),\\
	t_2(z)t_3(z)=4\left(z^4+z^2+1\right),
	\end{eqnarray}
which both remain positive for any value of $z$. And to conclude that the benchmark point $\xi^{Q,i}_{1,2,3,4}=1$ is fully consistent with the positivity bounds~\eqref{eq:selfbounds}.

\begin{figure} 
		\centering
		\textbf{Bounds on (4-Q) operators coefficients for $N_f=3$}\\\medskip
		\includegraphics[width=0.4\textwidth]{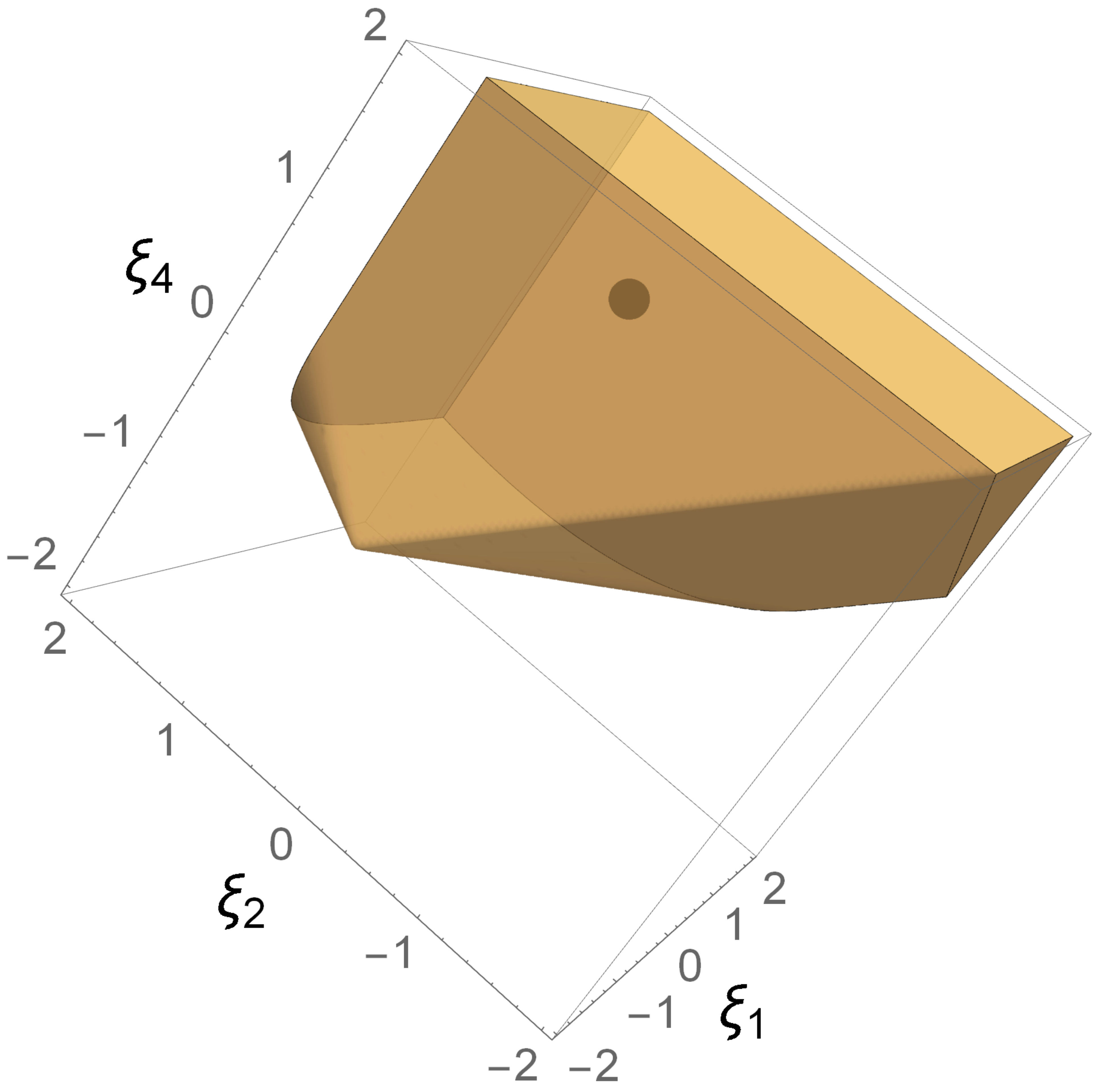}
		\caption{Plot showing in yellow the allowed region obtained for the (4-Q) (or, equivalently, (4-u)) operators with generic $\xi$ values. Using the scaling invariance of the bounds and since $\xi_3>0$, we have set $\xi_3=1$, and plot the remaining three independent coefficients. 
		The red dot indicates the \textit{natural} MFV benchmark point, $\xi_{1,2,3,4}=1$, that can be seen being inside the allowed region.}
		\label{fig:4Qprojections}
	\end{figure}

	We can then extend our analysis to generic values of the $\xi$ coefficients, as we did for $N_f=2$. The explicit computations are shown in Appendix~\ref{boundsGeneral}, while we report here only the resulting expression:
	\begin{align}
	\text{(4-Q) ($N_f$=3):  }
	\begin{cases}
	\bullet\  \xi^{Q,i}_3>0,\\
	\bullet\  \xi^{Q,i}_1+\xi^{Q,i}_3>0,\\
	\bullet\ \xi^{Q,i}_3+\xi^{Q,i}_4>0,\\
	\bullet\ \xi^{Q,i}_1+2 \xi^{Q,i}_2+2 \xi^{Q,i}_3+ 3 \xi^{Q,i}_4>0,\\
	\bullet\ \xi^{Q,i}_1+2 \xi^{Q,i}_2 +\xi^{Q,i}_3+2 \xi^{Q,i}_4 >0,\\
	\bullet\ \left( (\xi^{Q,i}_2-\xi^{Q,i}_4)^2-4 \xi^{Q,i}_3 \xi^{Q,i}_2\right)<4 \xi^{Q,i}_1 \left(\xi^{Q,i}_4 +\xi^{Q,i}_3\right)\\
	\qquad\qquad \text{ or } \left(\xi^{Q,i}_4-\xi^{Q,i}_2+2 \xi^{Q,i}_3\right) \left(\xi^{Q,i}_2+3 \xi^{Q,i}_4 +2 \xi^{Q,i}_3\right)>0.
	\end{cases}
	\label{eq:4uN3bounds}
	\end{align} 
	Those constraints are illustrated in Fig.~\ref{fig:4Qprojections}. 
	
%%%%%%%%%%%%%%%%%%%%%%%%%%%%%%%%%%%%%%%%%%%%%%%%%%%%%%%%%%%%%%%%%%%%%%%%%%%%%%%%%%%%%%%	
	\subsection{(4-u), (4-d) self-quartic and cross-quartic operators}
	Here we continue the discussion for the remaining operators. We will see that, in most cases, we have already done most of the work that was needed, and the bounds for the coefficients of these operators can simply be obtained by taking appropriate limits of the ones in Eqs.~\eqref{eq:4uN2bounds} and~\eqref{eq:4uN3bounds}, or by carefully looking at the order in which the indices are summed.
	
	We start by going through the remaining self-quartic operators, and then address the cross-quartic ones. 
	\paragraph{(4-u) operators}
	\label{sec:4Qand4u}
	As we have seen, the bounds we have found until now for the coefficients of the (4-Q) operators have turned out to depend exclusively on the eigenvalues of the matrix $M$. This is the case for the (4-u) operators, too, with the exchange $M\to\tilde{ M}$. However, the eigenvalues of these matrices coincide. This is clearly a basis-independent statement, but it can be easily seen in the basis in Eq.~\eqref{eq:interactionbasis2}, where they coincide. Thus, the resulting bounds are the same for the (4-u) operators as for the (4-Q) ones, with the simple replacement $\xi^{Q,i}_A\to\xi^{u,i}_A$ in Eqs.~\eqref{eq:4uN2bounds} and~\eqref{eq:4uN3bounds}. Similarly to what we did for the $(4-Q)$ case, we can, in the $N_f=2$ case, exploit the redundancy of $(\tilde{M}_{mq}\delta_{pn}+\delta_{mq}\tilde{M}_{pn})$ to remove two $\xi_4^{u,i}$ coefficients.
	
\begin{figure} 
		\centering
		\textbf{Bounds on (4-d) operators coefficients for $N_f=2$}\\\medskip
		\includegraphics[width=0.5\textwidth]{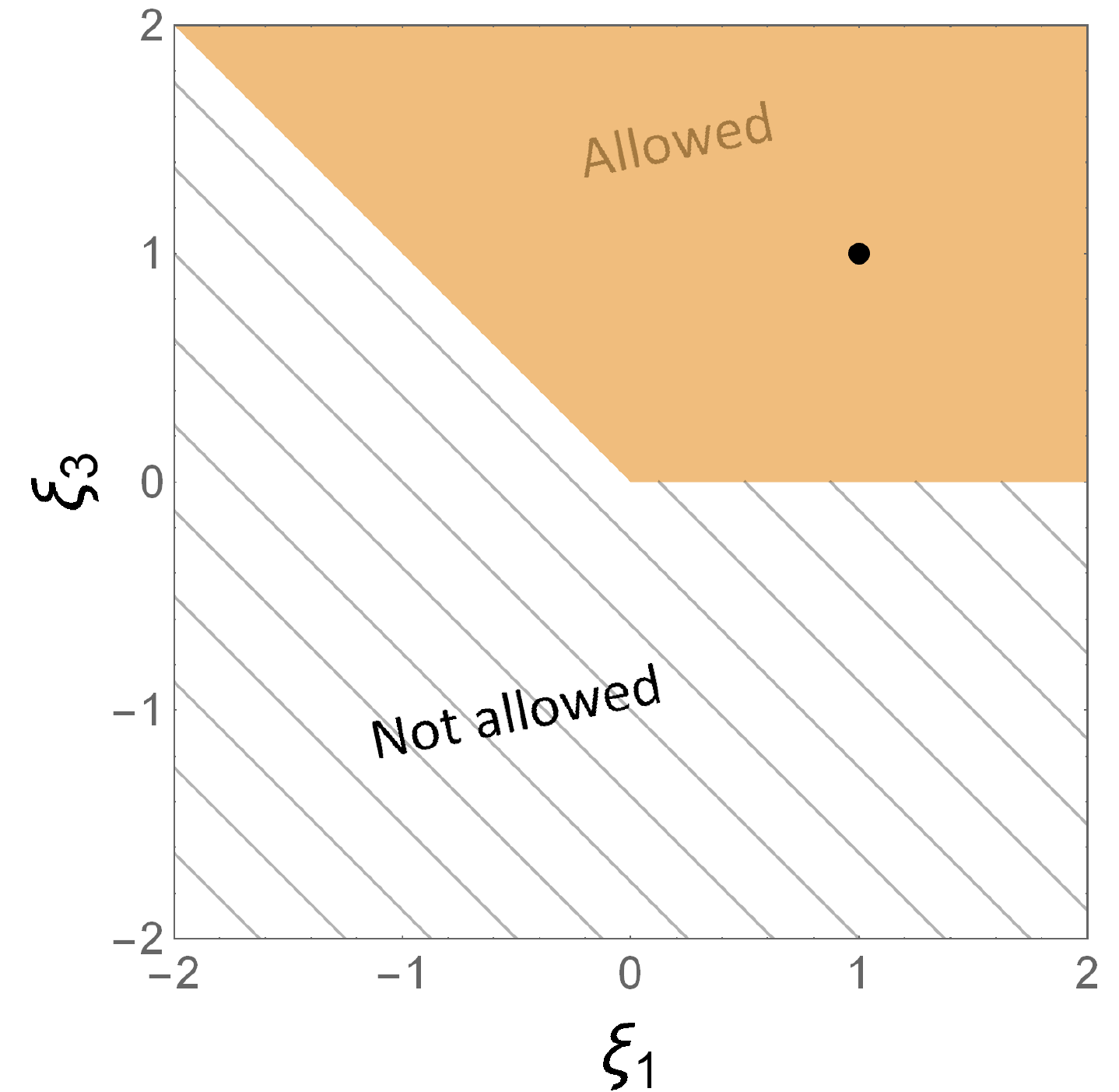}
		\caption{Plot showing in yellow the allowed parameter space for $\xi^{d,i}_1$ and $\xi^{d,i}_3$, $i=1,2$. The black dot indicates the \textit{natural} MFV benchmark point, $\xi_{1,3}=1$, that can be seen being inside the allowed region.}
		\label{fig:4d}
	\end{figure}

	An exception where the simple exchange $\xi^{Q,i}_A\to\xi^{u,i}_A$ does not work is the case studied in Section~\ref{sec:nf2revisisted}. Indeed, in this setting the two matrices differ, and in particular $\tilde{ M}=0$. The bounds are then retrieved in this case by sending $y_c\to0$ in Eq.~\eqref{eq:4uN2bounds}. This gives simply 
		\begin{align}
	\text{(4-u) ($N_f$=2 \text{revisited}):  }
	\begin{cases}
	\xi^{u,i}_3>0,\\
	\xi^{u,i}_1+\xi^{u,i}_3>0.\\ 
	\end{cases}
	\label{eq:4uN2boundsrevisited}
	\end{align}
	\paragraph{(4-d) operators}
	As anticipated, the bounds here can be obtained by applying a formal limit to the ones we have already. Indeed, we see that by \footnote{This is just a trick to get to the result, so in particular one does not need to worry about spoiling the Yukawa hierarchy that led to the approximation at the beginning of Section~\ref{sec:4Qand4u}.} sending $\tilde{M}\to0$ or $M\to0$ in the first two lines of Table~\ref{tab:MFV} respectively, we retrieve the tensor structure associated to the (4-d) operators. Then, the $N_f=2$ and $N_f=3$ cases produce the same bounds, the ones we already saw in Eq.~\eqref {eq:4uN2boundsrevisited}:
	\begin{align}
	\text{(4-d):   }
	\begin{cases}
	\xi^{d,i}_3>0,\\
	\xi^{d,i}_1+\xi^{d,i}_3>0.
	\end{cases}
	\end{align}
Figure~\ref{fig:4d} shows a plot of the corresponding allowed region. Again, we can note that the \textit{natural} MFV benchmark point, $\xi^{d,i}_{1,3}=1$, is compatible with the positivity constraints.
	\paragraph{Cross-quartic operators}
	All the bounds on the cross-quartic operators give in fact much less information than the ones on the self-quartic ones, in our MFV setting. This is due to the fact that, looking at the index disposition in $a_{mnpq}$ for the (2-d)(2-Q) and (2-d)(2-u) operators, we see that the objects in the l.h.s of Eq.~\eqref{eq:crossbounds} are essentially linear combinations of products of the norms of $\alpha$ and $\beta$, meaning terms of the form $\alpha_m A_{mq}\alpha_q^*\beta_n B_{np} \beta_p^*$, where one between $A_{mn}$ and $B_{mn}$ is a $\delta_{mn}$, while the other is either a $\delta$, or $M$, or $\tilde{M}$. Then, since all three of these matrices are (semi-)positive definite, taking all $\xi=1$ means that Eq.~\eqref{eq:crossbounds} turns into a sum of positive terms and the bounds are trivially satisfied. 
	
	The (2-u)(2-Q) case, however, has an additional term $\alpha_m\alpha^*_q\beta_n\beta^*_p(Y_u)_{nq}(Y^\dagger_u)_{mp}$. This is nothing but the modulus squared of $\beta_n Y_{u,nq} \alpha^*_q$, which is then also positive. Moreover, we can rest assured that, as long as we pick only positive values for the $\xi$ coefficients, the bounds will be fulfilled, so we can definitely find acceptable $\order{1}$ values for them, independently on the renormalizable Lagrangian parameters. This kills any hope of bounding them through these operators. 
	To get a full picture, one can here, too, allow for generic $\xi$ values, whose shape is dictated by the Yukawa couplings. Results as well as specific plots are shown in Appendix~\ref{boundsGeneral}.

Thus, in the end  we have proven that the \textit{natural} MFV benchmark point where all the flavor-blind factors of the MFV expansion  are chosen to be one, the positivity constraints of Ref.~\cite{flavorconstr} are all satisfied.

Moreover, since the constraints themselves are unchanged if multiplied by a positive factor, what we have showed is actually that any configurations where the $\xi$ coefficients are degenerate and positive are compatible with the positivity requirements.  Because of this scaling invariance, and of the particular shape of the constraints, when considering generic values for the $\xi$ coefficients, we can always rescale one of them to be $1$ or $-1$, once per every type of operators. Taking this into account, we see that, for physical values of the parameters, at least in the $N_f=3$ case, the parameter space spanned by the $\xi$ coefficients  is at least cut by a factor of two. Since there are a total of 14 independent types of operators under consideration, the overall allowed region is at least $2^{14}$ times smaller than the one with no positivity restriction\footnote{This has obviously to be understood as the result of a limit, meaning that if the space of parameters is restricted to a box of volume $\mathcal{V}$, then the allowed region has a volume $\sim 2^{-14}\mathcal{V}$, where $\mathcal{V}$ is eventually sent to infinity.}.
%%%%%%%%%%%%%%%%%%%%%%%%%%%%%%%%%%%%%%%%%%%%%%%%%%%%%%%%%%%%%%%%%%%%%%%%%%%%%%%%%%%%%%%	

\section{Discussion and Conclusions}
	\label{sec:conclusions}

In Ref.~\cite{flavorconstr}, using arguments that rely on the analyticity and unitarity of the theory in the UV, the authors obtained positivity constraints on the coefficients of dimension-8  operators with 4~fermions. Starting from that result, we showed that Minimal Flavor Violation, perhaps the simplest way to generalize the Standard Model flavor structure to higher dimensional operators, can be made consistent with those positivity constraints. To show this, we have first identified the bounds on the (flavor-blind) parameters that control the MFV expansion of the EFT coefficients. 
Such bounds are obtained after we disentangle the physical parameters from other parameters describing initial and final states of the $2\to2$ scattering processes. We have shown how this can be done in the specific case where the scattered states contain non-trivial flavor structure. This allowed us to find bounds on the various coefficients that parametrize the dimension-8 operators under the MFV assumption. In the space spanned by these coefficients, the positivity constraints become in general non-linear. This is a general feature that the linear nature of the positivity bounds exhibited in the simplest cases %is gradually lost as the scattered initial and final states  are dressed with less and less trivial structure under Lorentz, gauge, or global symmetries. Such a feature was already noticed in Ref.~\cite{aQGC} that studied the scattering of polarised states. 
(e.g. that of a single scalar or a single flavor of an uncharged Weyl fermion) is gradually lost when the number of degrees of freedom describing the scattered initial and final states is increased, for instance by considering non-trivial internal quantum numbers or several flavors, thereby increasing the number of dimension-8 operators to be studied.
	
	As a result, the \textit{natural} benchmark point, where all the flavor-blind parameters that enter the MFV expansion of EFT operators are degenerate and equal to unity, has proven to trivially satisfy the aforementioned positivity constraints. This is true independently of the parameters of the renormalizable SM Lagrangian, namely the fermion masses and the entries of the CKM matrix.  More generally, every setting where the flavor-blind parameters are degenerate is seen to be compatible with the positivity conditions.
Still, the positivity constraints are such as to reduce the full parameter space  by a factor of $2^{14}$.
	
Remarkably,  MFV is not restraining as to turn the positivity constraints on the dimension-8 operators into restrictions on the physical input parameters of the SM defined by the dimension-4 operators.  
	
	An immediate consequence is that, for flavor models which are less restricting than MFV, such as the so called $U(2)^5$ model~\cite{U(2)5,MFVandU2}, and which reduce to MFV for some values of the parameters, there exist at least an allowed region of the parameter space where the free coefficients are compatible with the bounds. 
	
It would also be interesting to derive (maybe) tighter constraints following the approach of Ref.~\cite{electronpositronscattering}
and considering the scattering of states that are no longer SM gauge eigenstates.  For instance, after suitable redefinitions accounting for the different operator basis used there, the bounds used in our work can be mapped into the ones in Eqs.~(7)--(10) of Ref.~\cite{electronpositronscattering}, while Eqs.~(11)--(12) are missing from our analysis\footnote{The precise signs of the bounds differ between Ref.~\cite{electronpositronscattering} and Ref.~\cite{flavorconstr}, which is likely to be due to different conventions. We chose conventions so that the signs are those given in Ref.~\cite{flavorconstr}. The fact that some bounds are missing in our analysis and in Ref.~\cite{flavorconstr} is anyway independent of these sign conventions.}. However, in Ref.~\cite{electronpositronscattering} only one flavor family is taken into account, whereas, due to the non-linearity of the additional bounds obtained, the $N_f\neq1$ case cannot be straightforwardly tackled using, e.g., the approach we outlined in Section~\ref{sec:studyingthebounds}. Therefore, we sticked for our analysis to the bounds of~\cite{flavorconstr}, and left the study of the most general scatterings to future work.

Finally, we point out that the analysis carried out in this paper immediately extends to the positivity constraints obtained in Ref.~\cite{dimension6bounds} on the dimension-6 4-Fermi operators. %in a MFV setup. 
Those bounds are derived under the assumption of an improved behavior of the scattering amplitude in the deep UV at energies above the EFT cutoff scale, behavior which is actually not encountered in simple UV-completions of the dimension-6 4-Fermi operators obtained by integrating out massive vectors.  Nevertheless, taking those bounds at face value and upon some linear redefinitions, %these said constraints 
they are of the same shape as those in Eqs.~\eqref{eq:selfbounds} and~\eqref{eq:crossbounds}, with $c_{mnpq}$ and $a_{mnpq}$ substituted by suitable combinations of their dimension-6 counterparts. This implies that the techniques we have exposed in the present paper also allow to derive constraints on the dimension-6 coefficients in a MVF setup, similar to, e.g., Eq.~\eqref{eq:4uN3bounds}. We leave the explicit presentation of the bounds in the dimension-6 case, as well as an analysis of their phenomenological impact, for future work.

%%%%%%%%%%%%%%%%%%%%%%%%%%%%%%%%%%%%%%%%%%%%%%%%%%%%%%%%%%%%%%%%%%%%%%%%%%%%%%%%%%%%%%%	
\section*{Acknowledgments}

We acknowledge support by the Deutsche Forschungsgemeinschaft under Germany's Excellence Strategy  EXC 2121 ``Quantum Universe" - 390833306. 	We thank M.~Montull for early discussions on positivity bounds and for comments on the draft of this paper, and J.~Gu with insightful discussions on the phenomenological relevance of the positivity constraints. We also thanks J.~Louis for stimulating conversations in the early stages of this work.
	
%%%%%%%%%%%%%%%%%%%%%%%%%%%%%%%%%%%%%%%%%%%%%%%%%%%%%%%%%%%%%%%%%%%%%%%%%%%%%%%%%%%%%%%	
\appendix 	
%%%%%%%%%%%%%%%%%%%%%%%%%%%%%%%%%%%%%%%%%%%%%%%%%%%%%%%%%%%%%%%%%%%%%%%%%%%%%%%%%%%%%%%%	
%\section{Group theoretical perspective}\label{sec:grouptheory}

%%%%%%%%%%%%%%%%%%%%%%%%%%%%%%%%%%%%%%%%%%%%%%%%%%%%%%%%%%%%%%%%%%%%%%%%%%%%%%%%%%%%%%%	
\section{Flavor-violation and large Yukawas}\label{sec:grouptheoryLargeTop}

	Our MFV expansion needs a little more justification, in particular concerning the resummation of the top-Yukawa $y_t$. Indeed, while $Y_d$ has eigenvalues $\ll 1$, and thus allows for the approximation we explained in Section~\ref{sec:MFVansatzSection}, this is not true for $Y_u$, whose biggest eigenvalue is $y_t\sim 1$. Consequently there is in principle no clear expansion in powers of $Y_u$ as long as we keep all the coefficients in the operator expansion of $\order{1}$. However, in Ref.~\cite{MFVreview}, for example, it is stated that any $\order{1}$ term in the (4-Q) case has to be of the form $\left(Y_uY^\dagger_u\right)^n$. In the basis of Eq.~\eqref{eq:interactionbasis2}, this reads:
\be
	\left(Y_uY^\dagger_u\right)^n_{ij}\sim y_t^{2n} (V_\textit{CKM}^*)_{3i}(V_\textit{CKM})_{3j} \ ,
	\label{flavorViolatingStructure}
\ee
	while the structure for the (4-u) case similarly reads $\left(Y^\dagger_uY_u\right)^n_{ij}\sim y_t^{2n} \delta_{3i}\delta_{3j}$. Consequently, considering more Yukawa matrices does not change the flavor structure of the couplings, but simply demands to resum the powers of $y_t$. We dwell a bit on the details of this conclusion below.
	
%%%%%%%%%%%%%%%%%%%%%%%%%%%%%%%%%%%%%%%%%%%%%%%%%%%%%%%%%%%%%%%%%%%%%%%%%%%%%%%%%%%%%%%	
\subsection{Group theory argument}
	
	The restriction to Eq.~\eqref{flavorViolatingStructure} can be justified like this: in $SU(3)$, the invariant tensors are $\epsilon_{abc}$, $\epsilon^{abc}$ and $\delta^a_b$. Imagine we want to build a contribution to the (4-Q) case using $n$ powers of $Y_u$ and $m$ powers of $Y_u^\dagger$. Then, since $\bar{Q}Q\bar{Q}Q$ is a singlet under $SU(3)_u$, we need to contract all of the $SU(3)_u$ indices of the various $Y_u$ and $Y_u^\dagger$ using $\epsilon$'s or $\delta$'s. If we contract the indices of three $Y_u$'s using an $\epsilon$, then for this product not to vanish, the $SU(3)_q$ indices of those matrices need to be fully antisymmetrized as well. However, this is but a singlet under $SU(3)_Q\otimes SU(3)_u$, and contributes as a redefinition of the coefficient of one operator of order $(Y_u)^{n-3}(Y_u^\dagger)^m$. Similarly for $Y_U^\dagger$. Then, we can only consistently use $\delta^a_b$ to contract the $SU(3)_u$ indices to hope to build a new structure. This proves that the building block is actually $Y_uY_u^\dagger$ which is a $\bar{\mathbf{3}}\otimes\mathbf{3}$ of $SU(3)_Q$, and we have to take $n=m$. Then, we want eventually to contract the remaining indices with those from $\bar{Q}Q\bar{Q}Q$. We want to prove that this, too, can be done exclusively with $\delta$'s. Indeed, suppose we want to employ $\epsilon_{abc}$. Similarly to what happened before, if we use it to contract three upstairs indices coming from three copies of $Y_uY_u^\dagger$, symmetry imposes that the downstairs indices are antisymmetrized, too, giving rise to an uninteresting singlet. If we contract it with two upstairs indices from two $Y_uY_u^\dagger$ and one from a $Q$, then the two downstairs indices form the two $Y_uY_u^\dagger$ need to be antisymmetrized with an $\epsilon^{abc}$. However the product of two epsilon is but a sum of products of deltas. Similarly, if we contract $\epsilon_{abc}$ with the two fundamental indices of the two $Q$, then the antifundamental ones of the two $\bar{Q}$ need to be contracted with an $\epsilon^{abc}$, again giving products of $\delta$'s. Thus a series in $\left(Y^\dagger_uY_u\right)^n$ includes all allowed contractions.
	Similar reasoning holds for the (4-u) case, with the exchange of $SU(3)_Q$ and $SU(3)_u$ indices, and $Y_uY_u^\dagger\to Y_u^\dagger Y_u$. 
	For the (4-d) operators, there are simply no possible non-trivial insertions of $Y_u$. (2-d)(2-Q) and (2-d)(2-u) are similar to the former cases, with the exception that there are now only one index in the fundamental and one in the antifundamental of $SU(3)_Q$ in the structure coming from the spinors.
	
	Finally, (2-Q)(2-u) requires a bit of attention. Again, contracting three $Y_u$'s or three $Y_u^\dagger$'s with an epsilon tensor eventually produces a $SU(3)_Q\otimes SU(3)_u$ singlet. However, the spinor structure provides an index in the fundamental and one in the antifundamental for both $SU(3)_u$ and $SU(3)_Q$. So we can contract two $Y_u$ or two $Y_u^\dagger$ with one index coming from the spinors. We cannot use an upstairs and a downstairs epsilon from the same group as that would reduce to sum of products of deltas. To sum up, we can use two epsilon tensors, one for each group, each with one index contracted to one coming from the spinor structure. Let us take the first one to be a $SU(3)_u$ $\epsilon^{abc}$. Its remaining two indices can only be contracted with two downstairs $SU(3)_u$ indices from two $Y_u$'s. Then, the $SU(3)_Q$ fundamental indices that these two matrices carry have to be antisymmetrized. This has to be done with the only remaining possible epsilon tensor, giving a structure as:
	\begin{align}
	(\bar{u}u)_{\bar{u}_1 u_1} (\bar{Q}Q)_{\bar{q}_1 q_1}(Y_u)_{q_2 \bar{u}_2}(Y_u)_{q_3 \bar{u}_3}\epsilon_{\bar{u}_1\bar{u}_2\bar{u}_3}\epsilon_{q_1q_2q_3}(Y_u^\dagger)_{u_1 \bar{q}_1}
	\label{eq:nontrivialstr}
	\end{align}
The other case, i.e., picking a $SU(3)_u$ $\epsilon_{abc}$, gives the hermitian conjugate of~\eqref{eq:nontrivialstr}.
However, this term is subleading, as it can be immediately seen by plugging the leading contribution $(Y_u)_{q \bar{u}}\sim y_t \delta_{3 \bar{u}}\delta_{3q}$. Moreover, it is not $U(1)_u$ invariant.
	In conclusion, only $\delta$ factors can be used to contract indices consistently. This forces the operators to be of the form already contained in the (2-u)(2-Q) line of Table~\ref{tab:MFV}, times an arbitrary number of $Y^\dagger_uY_u$ of $Y_uY_u^\dagger$, suitably contracted in. The latter can anyhow be reabsorbed in a redefinition of the overall $\rho$ coefficients, as we show in the next section.
	
%%%%%%%%%%%%%%%%%%%%%%%%%%%%%%%%%%%%%%%%%%%%%%%%%%%%%%%%%%%%%%%%%%%%%%%%%%%%%%%%%%%%%%%	
\subsection{Non-linear realization}
	
One can also phrase the argument in favor of the single flavor-violating structure in Eq.~\eqref{flavorViolatingStructure} in a non-linear language. Indeed, when $y_t$ is ${\cal O}(1)$, the necessary resummation of the expansion in powers of $y_t$ means that the flavor group is non-linearly realized~\cite{Feldmann:2008ja,Kagan:2009bn}. In the basis~\eqref{eq:interactionbasis2} where the up-type Yukawa is diagonal, the EFT is an expansion around the vev
\be
\langle Y_u\rangle =\bmat0&0&0\\0&0&0\\0&0&y_t\emat \ ,
\ee
breaking $U(3)_Q\times U(3)_u$ down to $U(2)_Q\times U(2)_u \times U(1)_3$. Following Ref.~\cite{Kagan:2009bn}, the building blocks for the EFT are found as follows. We first identify the Goldstone modes,
\be
Y_u=e^{i\hat\rho_Q}\bmat\phi_u&0\\0&y_t\emat e^{-i\hat\rho_u} \ , \text{ with } 
\hat\rho_i=\bmat 0&\rho_i\\\rho_i^\dagger&\theta_i\emat \ ,
\ee
with $\rho_i$ a complex 2-vector and $\theta_Q=-\theta_u\equiv\theta$, with $\theta$ a real field. The fields transform as
\be
e^{i\hat\rho_i}\rightarrow V_i e^{i\hat\rho_i} U^\dagger_i(\hat\rho_i,V_i) \ , \quad \bmat\phi_u&0\\0&y_t\emat \rightarrow U_Q(\hat\rho_Q,V_Q)\bmat\phi_u&0\\0&y_t\emat U^\dagger_u(\hat\rho_u,V_u) \ ,
\ee
where $U_i$ are functions of $V_i$ and of the Goldstones that belong to $U(2)_i\times U(1)_3$:
\be
U_i=\bmat U_i^{2\times 2}&0\\0&e^{i\phi_3}\emat \ .
\ee
We can also dress the down-type Yukawa matrix as well as some of the quark fields to obtain fields that transform as linear representations of $U(2)_Q\times U(2)_u\times U(1)_3$ under the full flavor group:
\be
\tilde Y_d\equiv e^{-i\hat\rho_Q}Y_d \ , \quad \tilde Q\equiv e^{-i\hat\rho_Q}Q \ ,\quad \tilde u\equiv e^{-i\hat\rho_u}u \ .
\ee
The new fields can be split as follows:
\be
\tilde Y_d=\bmat \phi_d\\ \phi'^\dagger_d \emat \ , \quad \tilde Q=\bmat \tilde Q^{(2)}\\\tilde t_L\emat  \ , \quad \tilde u=\bmat \tilde u^{(2)}\\\tilde t_R\emat \ ,
\ee
where $\phi_d$ is a $2\times 3$ matrix, $\phi'_d$ a 3-vector, $\tilde Q^{(2)}$ a doublet of $U(2)_Q$, $\tilde t_L$ a singlet, and similarly for $\tilde  u$. The components transform as
\begin{align}
&\phi_d \rightarrow U_Q^{2\times 2}\phi_d V_d^\dagger \ , \quad \phi'_d \rightarrow e^{-i\phi_3}V_d\phi'_d \ , \quad \tilde Q^{(2)}\rightarrow U_Q^{2\times 2} \tilde Q^{(2)} \ , \\
&\tilde t_L\rightarrow e^{i\phi_3}\tilde t_L \ , \quad \tilde u^{(2)}\rightarrow U_u^{2\times 2}\tilde u^{(2)} \ , \quad \tilde t_R\rightarrow e^{i\phi_3}\tilde t_R \ .
\end{align}
The fields above, together with $d_R,\phi_u,y_t$ and the invariance under $U(2)_Q\times U(2)_u \times U(1)_3\times U(3)_d (\times U(3)_L \times U(3)_e)$, are the building blocks for the EFT. One should in principle also use the covariant derivatives obtained from the Maurer-Cartan form
\be
e^{-i\hat\rho_Q}\partial_\mu e^{i\hat\rho_Q}
\ee
but they are identically zero when we freeze the Yukawa spurions to their background values. Summing up, the different fields and their representations are:
\begin{table}[H]
\centering
\begin{tabular}{c|c|c|c|c}
			& $U(2)_Q$ & $U(2)_u$& $U(1)_3$ & $U(3)_d$\\\hline
\multicolumn{5}{c}{Fermions}\\\hline
$\tilde Q^{(2)}$ &$\mathbf{2}$ & $ \mathbf{1}$ & $0$ &$\mathbf{1}$  \\
$\tilde t_L$ &$\mathbf{1}$ & $ \mathbf{1}$ & $+1$ &$\mathbf{1}$  \\
$\tilde u^{(2)}$ &$\mathbf{1}$ & $ \mathbf{2}$ & $0$ &$\mathbf{1}$  \\
$\tilde t_R$ &$\mathbf{1}$ & $ \mathbf{1}$ & $+1$ &$\mathbf{1}$  \\
$\tilde d$ &$\mathbf{1}$ & $ \mathbf{1}$ & $0$ &$\mathbf{3}$  \\
\hline
\multicolumn{5}{c}{Spurions}\\\hline
			$\phi_u$ &$\mathbf{2}$ & $ \mathbf{\bar{2}}$ & $0$ &$\mathbf{1}$  \\
			$y_t$ &$\mathbf{1}$ & $ \mathbf{1}$ & $0$ &$\mathbf{1}$  \\
			$\phi_d$ &$\mathbf{2}$ & $ \mathbf{1}$&0 & $\mathbf{\bar{3}}$  \\
			$\phi'_d$ &$\mathbf{1}$ & $ \mathbf{1}$&-1 & $\mathbf{3}$  \\

\end{tabular}
\end{table}
The background values of the spurions are obtained from $Y_u=\text{diag}(y_u,y_c,y_t)$, $Y_d=V_\textit{CKM}\text{diag}(y_d,y_s,y_b)$.
We see that all spurions but $y_t$ are small, so that there exists an expansion in terms of small Yukawas and CKM elements. Every EFT term constructed from $\phi_u,\phi_d,\phi'_d$ up to a given order is then completed by multiplying it by an arbitrary function of $y_t$. In particular, the approximation we have been discussing in this appendix is the one where all $y$s but $y_t$ are zero. At this order, the only fermion bilinears that can enter the dimension-8 coefficients in Table~\ref{tab:operators} in a flavor-invariant way are
\be
\overline {\tilde Q^{(2)}}\gamma^\mu \tilde Q^{(2)} \ , \quad \overline {\tilde t_L}\gamma^\mu \tilde t_L \ , \quad \overline {\tilde u^{(2)}}\gamma^\mu \tilde u^{(2)} \ , \quad \overline {\tilde t_R}\gamma^\mu \tilde t_R \ , \quad \overline {\tilde d}\gamma^\mu \tilde d \ .
\label{LONonLinear}
\ee
When we freeze the spurions to their background values, the Goldstone fields $\hat\rho$ are zero in the basis where $Y_u$ is diagonal, one can simply remove the tildes in the expression above and rename $\tilde t_L=Q_3,\tilde t_R=u_3$. In the basis where $Y_u=V_\textit{CKM}^\dagger\text{diag}(y_u,y_c,y_t)$ and $Y_d=\text{diag}(y_d,y_s,y_b)$, Eq.~\eqref{LONonLinear} becomes
\be
\bead
\left[\delta_{ij}-\left(V_\textit{CKM}^*\right)_{3i}\left(V_\textit{CKM}\right)_{3j}\right]\overline {Q_{i}}\gamma^\mu Q_{j} \ ,& \quad \left(V_\textit{CKM}^*\right)_{3i}\left(V_\textit{CKM}\right)_{3j}\overline {Q_{i}}\gamma^\mu Q_{j} \ ,\\
\overline {u^{(2)}}\gamma^\mu u^{(2)} \ , \quad \overline {u_3}&\gamma^\mu u_3 \ , \quad \overline {d}\gamma^\mu d \ .
\eead
\ee
This is consistent with Eq.~\eqref{flavorViolatingStructure}: the two terms in the first line can be combined to reconstruct $\delta_{ij}\overline {Q_{i}}\gamma^\mu Q_{j}$ and $\left(V_\textit{CKM}^*\right)_{3i}\left(V_\textit{CKM}\right)_{3j}\overline {Q_{i}}\gamma^\mu Q_{j}$, which are the flavor structures that are obtained from $\left(Y_uY^\dagger_u\right)^n$. The $u$-quark terms in the second line can be combined to reconstruct $\delta_{ij}\overline {u_{i}}\gamma^\mu u_{j}$ and $\delta_{3i}\delta_{3j}\overline {u_{i}}\gamma^\mu u_{j}$, which are the flavor structures that are obtained from $\left(Y^\dagger_uY_u\right)^n$. The $d$-quark terms are flavor diagonal, as they should at order $Y_d^0$.

%%%%%%%%%%%%%%%%%%%%%%%%%%%%%%%%%%%%%%%%%%%%%%%%%%%%%%%%%%%%%%%%%%%%%%%%%%%%%%%%%%%%%%%
\section{Redundancy of the $\rho_4$ structure in $N_f=2$}\label{sec:grouptheoryrho4}

Here we provide a proof showing that, in $SU(2)$, $\bar{u}_m u_n\bar{u}_p u_q(\tilde{ M}_{mq}\delta_{pn}+\delta_{mq}\tilde{ M}_{pn})$ is redundant with respect to the other structures contained in the first line of Table~\ref{tab:MFV}, and the corresponding coefficient, which we called $\rho_4$, can be reabsorbed in the definitions of the remaining three. An analogous discussion can be done for $\bar{u}_m u_n\bar{u}_p u_q(\tilde{ M}_{mq}\delta_{pn}+\delta_{mq}\tilde{ M}_{pn})$ and the second line of Table~\ref{tab:MFV}. First of all, define $X_{mn}\equiv \bar{u}_mu_n$. $SU(2)$ does not distinguish between fundamental and antifundamental indices, and all summations need to be performed with $\epsl{ab}$, the only invariant tensor. Then
\begin{align}
X_{ij}X_{kl}\epsl{jk}\epsl{lm}\epsl{in}\tilde{ M}_{nm}&=\left(X_{\{ij\}}+\frac{1}{2}X_{ab}\epsl{ab}\epsl{ij}\right)\left(X_{\{kl\}}+\frac{1}{2}X_{cd}\epsl{cd}\epsl{kl}\right)\epsl{jk}\epsl{lm}\epsl{in}\tilde{ M}_{nm}=\nonumber\\
&=X_{\{ij\}}X_{\{kl\}}\epsl{jk}\epsl{lm}\epsl{in}\tilde{ M}_{nm}+X_{ab}\epsl{ab}\epsl{nk}\epsl{lm}X_{\{kl\}}\tilde{ M}_{nm}+\nonumber\\
&\quad-\frac{1}{4}\epsl{ab}X_{ab}\epsl{cd}X_{cd}\epsl{nm}\tilde{ M}_{nm}.
\label{eq:eq1}
\end{align}
The second piece can be absorbed by a shift in $\rho_1$, while the third one with a shift on $\rho_2$. The first piece, instead, splits as:
\begin{align}
X_{\{ij\}}X_{\{kl\}}\epsl{jk}\epsl{lm}\epsl{in}\tilde{ M}_{nm}&=X_{\{ij\}}X_{\{kl\}}\epsl{jk}\epsl{lm}\epsl{in}\left(\tilde{ M}_{\{nm\}}+\frac{1}{2}\epsl{nm}\epsl{ef}\tilde{ M}_{ef}\right)=\nonumber\\
&=X_{\{ij\}}X_{\{kl\}}\tilde{ M}_{\{nm\}}\epsl{jk}\epsl{lm}\epsl{in}+\frac{1}{2}X_{\{ij\}}X_{\{kl\}}\epsl{jk}\epsl{il}\epsl{ef}M_{ef}.
\end{align}
Here, the second term is reabsorbed through a shift on $\rho_3$, while the first one vanishes:
\begin{align}
X_{\{ij\}}X_{\{kl\}}M_{\{nm\}}\epsl{jk}\epsl{lm}\epsl{in}&\overset{j\leftrightarrow k}{=}-X_{\{ij\}}X_{\{kl\}}M_{\{nm\}}\epsl{kj}\epsl{lm}\epsl{in}\overset{(ij)\leftrightarrow (kl)}{=}\nonumber\\
&=-X_{\{ij\}}X_{\{kl\}}M_{\{nm\}}\epsl{il}\epsl{jm}\epsl{kn}\overset{m\leftrightarrow n}{=}\nonumber\\
&=-X_{\{ij\}}X_{\{kl\}}M_{\{nm\}}\epsl{il}\epsl{jn}\epsl{km}\overset{i\leftrightarrow j}{=}\nonumber\\
&=-X_{\{ij\}}X_{\{kl\}}M_{\{nm\}}\epsl{jl}\epsl{in}\epsl{km}\overset{l\leftrightarrow k}{=}\nonumber\\
&=-X_{\{ij\}}X_{\{kl\}}M_{\{nm\}}\epsl{jk}\epsl{lm}\epsl{in}.
\end{align}
In conclusion, the examined term provides no new structure and can be set to 0.
	
%%%%%%%%%%%%%%%%%%%%%%%%%%%%%%%%%%%%%%%%%%%%%%%%%%%%%%%%%%%%%%%%%%%%%%%%%%%%%%%%%%%%%%%
\section{Contribution of other SMEFT operators}
\label{sec:dimension6operators}

One comment should also be made about the contributions of other SMEFT operators to the positivity bounds. Indeed, in the usual version of the latter that we consider in this paper, they are associated to the $s^2$ coefficient of a $2\to 2$ forward amplitude. By dimensional analysis only, one sees that such a $s^2$ growth can be obtained from a dimension-8 four-fermions contact term, as we considered in the main text, but also from the product of two coefficients, both of dimension-6, or one of dimension-5 and the other of dimension-7, or one of dimension-8 and at least one of dimension-4.

Let us discuss first the case of  dimension-6 operators. We mentioned already that dimension-6 four-fermions operators do not enter the positivity bounds, due to their softer UV behaviour when compared to dimension-8 operators. Nevertheless, there are operators at dimension-6 which, when combined together, contribute at tree level to the four-fermions amplitudes with a UV behaviour similar to that of dimension-8 four-fermions operators. A simple example is the following: consider the dimension-6 operator that couples the photon to the up-type right-handed quark current,
\begin{equation}
{\cal L} \supset \frac{c}{\Lambda^2}\partial_\mu J_\nu F^{\mu\nu} \ ,
\label{dimension6contributing}
\end{equation}
where $J_\mu=\overline{u}_m\gamma_\mu u_m$. This coupling generates a four-fermion amplitude whose s-channel component is depicted below and reads
\begin{center}
	\begin{tabular}{ll}
		\includegraphics[width=0.3\textwidth,valign=M]{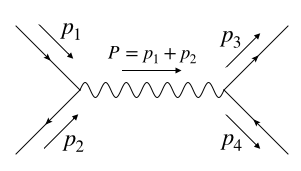}
		&
		\begin{minipage}{10cm}
			\begin{equation}
			\begin{aligned}
			{\cal A}=& \ i\frac{c^2}{\Lambda ^4}\overline{v_2}\gamma_\nu u_1P^2\left(\frac{\eta^{\nu\sigma}}{P^2}-(1-\xi)\frac{P^\nu P^\sigma}{(P^2)^2}\right)P^2\overline{u_3}\gamma_\sigma v_4\\
			\approx&\ i\frac{c^2}{\Lambda ^4}P^2\overline{v_2}\gamma_\mu u_1\overline{u_3}\gamma^\mu v_4 \ ,
			\end{aligned}
			\label{eq:dim6diagram}
			\end{equation}
		\end{minipage}
	\end{tabular}
\end{center}
where $u_i,v_i$ are (anti)particle polarizations and we used the fact that the particles are effectively massless at high energies, so that $\slashed{p}u(p),\slashed{p}v(p)\approx 0$. We would have obtained the same result with the four fermion interaction $\frac{c^2}{\Lambda ^4}\partial_\mu J_\nu \partial^\mu J^\nu$, which is of the kind subject to positivity bounds. Thus, $c^2$ should be added to the combination of dimension-8 coefficients that are constrained to be positive, modifying the bound. However, Eq.~\eqref{dimension6contributing} is a redundant operator that can consistently be ignored in the first place. Indeed, enforcing the photon equation of motion derived from Eq.~\eqref{dimension6contributing},
\begin{equation}
\partial^\mu F_{\mu\nu}-\frac{c}{\Lambda ^2}\Box J_\nu=0 \ ,
\end{equation}
we find that $\frac{c}{\Lambda ^2}\partial_\mu J_\nu F^{\mu\nu}=\frac{c^2}{\Lambda ^4}\partial_\mu J_\nu \partial^\mu J^\nu$, consistently with our previous analysis. Thus, we can set $c=0$ at no cost. Using the Warsaw basis for the dimension-6 SMEFT~\cite{Warsaw}, and considering the high-energy phase where all particles are massless and the electroweak symmetry unbroken, it is straightforward to check that there are no dimension-6 contributions to the four-fermion positivity bounds at tree level. This differs from the case of $2\to 2$ gauge bosons scattering, where non-redundant trilinear couplings enter the bounds at tree-level and strenghthen the bounds on dimension-8 operators \cite{vectorbosonscattering, aQGC,Yamashita:2020gtt}.

Similarly, contributions that could spoil our bounds come from combinations of dimension-5 and dimension-7, or of dimension-4 and dimension-8 terms. Since in the SMEFT the only dimension-5 term is the Weinberg operator $HHL_iL_j$, which does not contain quarks, the former possibility is irrelevant. 
Instead, combining one dimension-4 and one dimension-8 operator could give rise to a diagram with the same shape as the one in Eq.~\eqref{eq:dim6diagram} or an analogous one with an intermediate Higgs boson. The former could arise from combining a gauge interaction and a dimension-8 operator of the schematic form $\mathcal{O}\sim\partial^3X\bar{\psi}\psi$, with $X$ any of the SM field strengths, and the latter from combining a Yukawa interaction and an operator $\mathcal{O}\sim\partial^4 H \bar{\psi}\psi$.
However, in both cases the dimension-8 operator is proportional to some equations of motion~\cite{dimension8_1,dimension8_2}, and can thus be reabsorbed by redefining the coefficients of lower-dimensional operators, as just shown for the dimension-6 ones. 

Those conclusions can be reached in a number of ways, including Hilbert series techniques and on-shell methods. In conclusion, it is useful to have in mind that specifying the basis of operators is necessary when writing positivity bounds. 

%%%%%%%%%%%%%%%%%%%%%%%%%%%%%%%%%%%%%%%%%%%%%%%%%%%%%%%%%%%%%%%%%%%%%%%%%%%%%%%%%%%%%%%	
\section{Bounds for generic $\xi$ values}\label{boundsGeneral}
%%%%%%%%%%%%%%%%%%%%%%%%%%%%%%%%%%%%%%%%%%%%%%%%%%%%%%%%%%%%%%%%%%%%%%%%%%%%%%%%%%%%%%%	
\subsection{Self-quartic}

We report here the bounds that we get when we allow for generic values of the $\rho$ coefficients in the $c_{mnqp}$ and $a_{mnpq}$ tensors. The procedure is exactly the same as in the (4-Q) case with $N_f=2$, which was shown in Section~\ref{sec:4upnf2}. We keep explicit factors of $y_t^2$ to make it easier to identify the contributions of contracted Yukawa matrices, but we remind that they should be put to $1$ for consistency of the expansion.
	
Let us see for example how the (4-Q) case changes. Again, we impose the first eigenvalue of the matrix $C(\beta)$, as well as the sum and product of the remaining two, to be positive:
	\begin{align}
	t_1(z)&=\xi^{Q,i}_3+\xi^{Q,i}_3 y_t^2  z^2,\\
	t_2(z)+t_3(z)&=\xi^{Q,i}_1+2 \xi^{Q,i}_3+\xi^{Q,i}_4 y_t^2 +2 y_t^2  z^2 (\xi^{Q,i}_2+\xi^{Q,i}_4),\\
	t_2(z)t_3(z)&=(\xi^{Q,i}_1+\xi^{Q,i}_3) (\xi^{Q,i}_3+\xi^{Q,i}_4 y_t^2 )\nonumber\\
	&\hphantom{=}+ z^2y_t^2  (\xi^{Q,i}_2+\xi^{Q,i}_4) (y_t^2  (\xi^{Q,i}_4-\xi^{Q,i}_2)+2 \xi^{Q,i}_3)\nonumber\\
	&\hphantom{=}+z^4y_t^4 (\xi^{Q,i}_2+\xi^{Q,i}_4)^2.
	\end{align}
The first two are linear objects in $z^2$, which varies within the interval $[0,1]$. Thus, they are positive $\forall z$ if and only if their value at the boundaries of the interval is also positive. However, the third expression is a quadratic polynomial in $z^2$, which we can parametrize as $az^4+b x^2+c$. Then, as already seen, we require this polynomial to be positive at the boundaries and to satisfy $\Delta<0 \lor a<0 \lor b(b+2a)>0$.
	Putting everything together, and after some simplifications, we get the full set of conditions:
	\begin{align}
	\text{(4-Q):   }
	\begin{cases}
	\bullet\ \xi^{Q,i}_3>0,\\
	\bullet\ \xi^{Q,i}_1+\xi^{Q,i}_3>0,\\
	\bullet\ \xi^{Q,i}_3+\xi^{Q,i}_4 y_t^2>0,\\
	\bullet\ \xi^{Q,i}_1+2 \xi^{Q,i}_2 y_t^2+2 \xi^{Q,i}_3+ 3 \xi^{Q,i}_4 y_t^2>0,\\
	\bullet\ \xi^{Q,i}_1+2 \xi^{Q,i}_2y_t^2 +\xi^{Q,i}_3+2 \xi^{Q,i}_4 y_t^2 >0,\\
	\bullet\ y_t^2 \left(y_t^2 (\xi^{Q,i}_2-\xi^{Q,i}_4)^2-4 \xi^{Q,i}_3 \xi^{Q,i}_2\right)<4 \xi^{Q,i}_1 \left(\xi^{Q,i}_4 y_t^2 +\xi^{Q,i}_3\right)\\
	\qquad\qquad\textrm{or}\quad \left(y_t^2 (\xi^{Q,i}_4-\xi^{Q,i}_2) +2 \xi^{Q,i}_3\right) \left(y_t^2 (\xi^{Q,i}_2+3 \xi^{Q,i}_4) +2 \xi^{Q,i}_3\right)>0.
	\end{cases}
	\end{align} 
Again, since within our basis choice $M$ and $\tilde{M}$ are the same, identical results apply to the $\xi^{u,i}_k$, $k=1,2,3,4$.
	
%%%%%%%%%%%%%%%%%%%%%%%%%%%%%%%%%%%%%%%%%%%%%%%%%%%%%%%%%%%%%%%%%%%%%%%%%%%%%%%%%%%%%%%
\subsection{Cross-quartic operators}

\begin{figure} 
		\centering
		\textbf{Bounds on (2-u)(2-Q) operators coefficients for $N_f=3$}\\\medskip
		\includegraphics[width=0.5\textwidth]{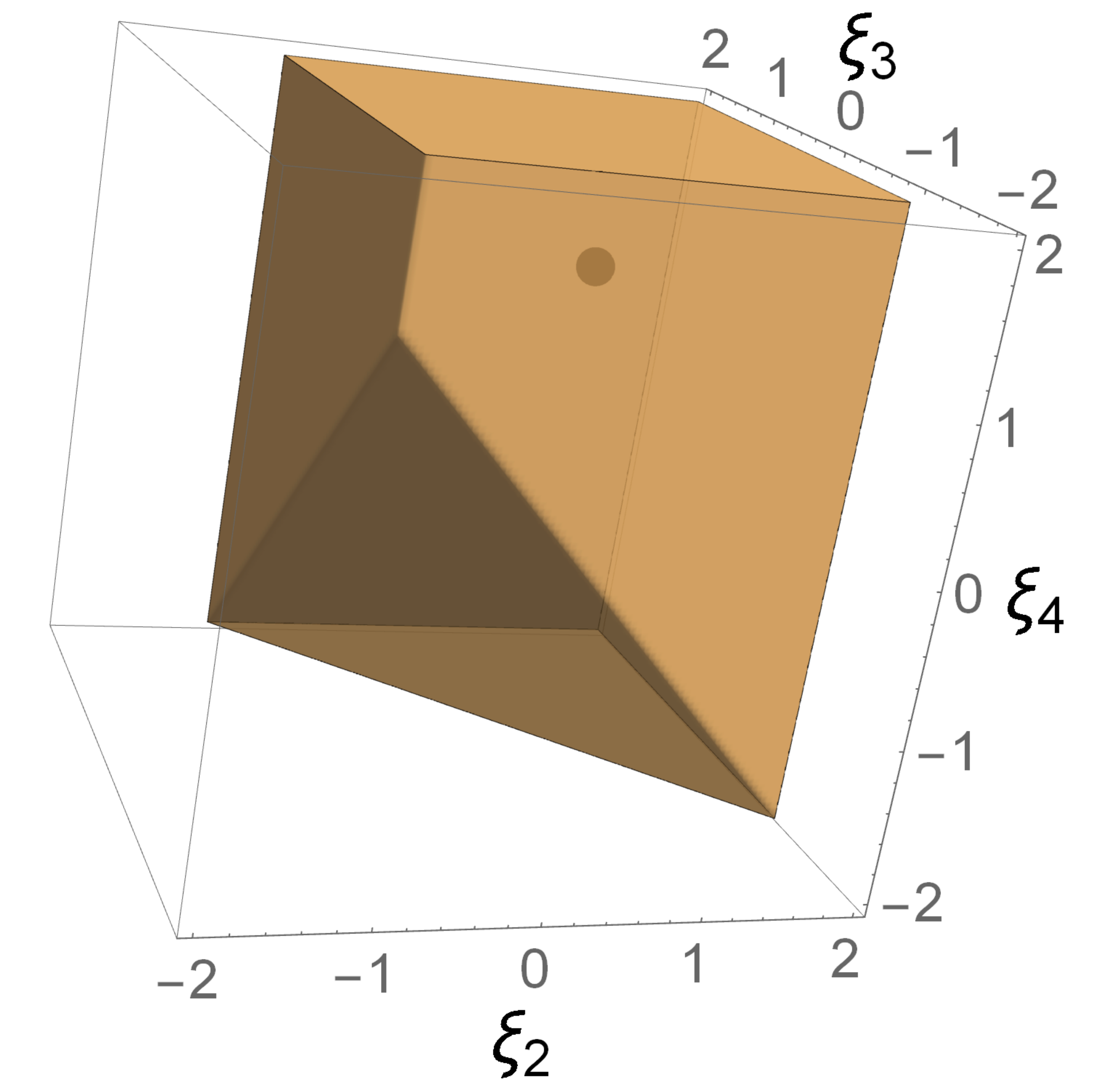}
		\caption{Plots showing in yellow the allowed region obtained for the (2-u)(2-Q) operators, with generic $\xi$ values. Here $\xi_1>0$, so we have rescaled it to 1, and we show the allowed region for the remaining three. $y_t$ is set to 1. The red dot indicates the \textit{naturak} MFV benchmark point, $\xi_{1,2,3,4}=1$, which can be seen being inside the allowed region.}
		\label{fig:4u4Qprojections}
	\end{figure}

The $C(\beta)$ matrix turns out to have two equal eigenvalues, and the conditions are simply:
	\begin{align}
	\text{(2-u)(2-Q):   }
	\begin{cases}
	\bullet\ \xi^{uQ,i}_1>0,\\
	\bullet\ \xi^{uQ,i}_1+\xi^{uQ,i}_2 y_t^2 >0,\\
	\bullet\ \xi^{uQ,i}_1+\xi^{uQ,i}_3 y_t^2 >0,\\
	\bullet\ \xi^{uQ,i}_1+(\xi^{uQ,i}_2+\xi^{uQ,i}_3+\xi^{uQ,i}_4)y_t^2>0.
	\end{cases}
	\end{align}
A plot of the results is reported in Fig.~\ref{fig:4u4Qprojections}.
	Here we also report the bounds for the (2-d)(2-Q) and the (2-d)(2-u) operators, which can both be computed directly or obtained from the previous ones with appropriate limits. Again, since $M$ and $\tilde{M}$ coincide, so do the bounds:
	\begin{align}
	\text{(2-d)(2-Q):   }
	\begin{cases}
	\bullet\ \xi^{dQ,i}_1>0,\\
	\bullet\ \xi^{dQ,i}_1+\xi^{dQ,i}_2 y_t^2 >0.
	\end{cases}
	\end{align}
	\begin{align}
	\text{(2-d)(2-u):   }
	\begin{cases}
	\bullet\ \xi^{du,i}_1>0,\\
	\bullet\ \xi^{du,i}_1+\xi^{du,i}_2 y_t^2 >0.
	\end{cases}
	\end{align}
%%%%%%%%%%%%%%%%%%%%%%%%%%%%%%%%%%%%%%%%%%%%%%%%%%%%%%%%%%%%%%%%%%%%%%%%%%%%%%%%%%%%%%%
\section{$\rho$ dependence of $\xi$ coefficients}
\label{sec:rhodependence}

\paragraph{Self-quartic}
	For the operators built out of (4-Q) fields, the coefficient looks like:
	\begin{align}
	c^{Q,i}_{mnpq}&=\rho^{Q,i}_1(\delta_{mn}\delta_{pq})+\rho^{Q,i}_2(M_{mn}\delta_{pq}+\delta_{mn}M_{pq})+\rho^{Q,i}_3(\delta_{mq}\delta_{pn})+\nonumber\\
	&+\rho^{Q,i}_4(M_{mq}\delta_{pn}+\delta_{mq}M_{pn})
	\end{align}
	The conditions coming from Eq. (12) of Ref.~\cite{flavorconstr} are:
	\begin{align}
	\alpha_m\alpha^*_q\beta_n\beta^*_p \left(c_{mnpq}^{Q,1}+\frac{1}{4}c_{mnpq}^{Q,2}+\frac{1}{3}c_{mnpq}^{Q,3}+\frac{1}{12}c_{mnpq}^{Q,4}\right)&>0,\nonumber\\
	\alpha_m\alpha^*_q\beta_n\beta^*_p \left(c_{mnpq}^{Q,2}+\frac{1}{3}c_{mnpq}^{Q,4}\right)&>0,\nonumber\\
	\alpha_m\alpha^*_q\beta_n\beta^*_p \left(c_{mnpq}^{Q,3}+\frac{1}{4}c_{mnpq}^{Q,4}\right)&>0,\nonumber\\
	\alpha_m\alpha^*_q\beta_n\beta^*_p c_{mnpq}^{Q,4}&>0.
	\end{align}
	Then, we can define the linearly transformed coefficients $\xi(\rho)$:
	\begin{align}
	&\begin{cases}
	\bullet\ \xi^{Q,1}_k&\equiv\rho^{Q,1}_k+\frac{1}{4}\rho^{Q,2}_k+ \frac{1}{3}\rho^{Q,3}_k+\frac{1}{12}\rho^{Q,4}_k,\\
	\bullet\ \xi^{Q,2}_k&\equiv\rho^{Q,2}_k+\frac{1}{3}\rho^{Q,4}_k,\\
	\bullet\ \xi^{Q,3}_k&\equiv\rho^{Q,3}_k+\frac{1}{4}\rho^{Q,4}_k,\\
	\bullet\ \xi^{Q,4}_k&\equiv\rho^{Q,4}_k,
	\end{cases}& k&=1,2,3,4,
	\end{align}
	to turn the bounds into 
	\begin{align}
	\alpha_m\alpha^*_q\beta_n\beta^*_p c(\xi)_{mnpq}^{Q,i}&>0 & i&=1,2,3,4,
	\end{align}
i.e., in the form of Eq.~\eqref{eq:selfbounds}.
	For the (4-d) operators, the bounds are the same as in the (4-u) case:
	\begin{align}
	\alpha_m\alpha^*_q\beta_n\beta^*_p \left(c_{mnpq}^{d,1}+\frac{1}{3}c_{mnpq}^{d,3}\right)&>0,\nonumber\\
	\alpha_m\alpha^*_q\beta_n\beta^*_p c_{mnpq}^{u,3}&>0.
	\end{align}
	Thus, with a redefinition identical to~\eqref{eq:upredefinition}, we recast them into the desired form in Eq.~\eqref{eq:selfbounds}.
	\paragraph{Cross-quartic}
	The bounds on the (2-u)(2-Q) operators are:
	\begin{align}
	\alpha_m\alpha^*_q\beta_n\beta^*_p\left(a^{uQ,1}_{mnpq}+\frac{1\pm 3}{12}a^{uQ,3}_{mnpq}\right)>0.
	\end{align}
	Upon defining:
	\begin{align}
	&\begin{cases}
	\xi^{uQ,1}_k&\equiv\rho^{uQ,1}_k+\frac{1}{3}\rho^{uQ,3}_k,\\
	\xi^{uQ,3}_k&\equiv\rho^{uQ,1}_k-\frac{1}{6}\rho^{uQ,3}_k,
	\end{cases}& k&=1,2,3,4,
	\end{align}
they are recast as in Eq.~\eqref{eq:crossbounds}.
The shape of the bounds on the remaining cross quartic operators is the same, therefore identical redefinitions
	\begin{align}
	&\begin{cases}
	\xi^{dQ,1}_k&\equiv\rho^{dQ,1}_k+\frac{1}{3}\rho^{dQ,3}_k, \\
	\xi^{dQ,3}_k&\equiv\rho^{dQ,1}_k-\frac{1}{6}\rho^{dQ,3}_k,  
	\end{cases} & k&=1,2,\\
	&\begin{cases}
	\xi^{du,1}_k&\equiv\rho^{du,1}_k+\frac{1}{3}\rho^{du,3}_k, \\
	\xi^{du,3}_k&\equiv\rho^{du,1}_k-\frac{1}{6}\rho^{du,3}_k, 
	\end{cases}& k&=1,2,
	\end{align}
are needed to recast the bounds in the form of Eq.~\eqref{eq:crossbounds}.

%%%%%%%%%%%%%%%%%%%%%%%%%%%%%%%%%
\bibliography{references}{}
\bibliographystyle{JHEP}

\end{document}